\documentclass[12pt,a4paper,fleqn]{article}
\usepackage{amsmath}
\usepackage{graphicx}
\usepackage{cite}
\usepackage{fixme}
\usepackage{multirow}

\usepackage[american]{babel}

\usepackage{dcolumn} 
\newcolumntype{d}{D{.}{.}{2}}
\newcolumntype{e}{D{.}{.}{3}}
\newcolumntype{f}{D{.}{.}{4}}

\usepackage{endfloat} 
\AtBeginDelayedFloats{\linespread{1.3}}

\begin{document}

\begin{center}

{\LARGE\bf
Optimized Unrestricted Kohn--Sham Potentials from \emph{Ab Initio} Spin Densities
}

{\large 
Katharina Boguslawski$^{\rm a}$,
Christoph R. Jacob$^{\rm b,}$\footnote{Corresponding Author; E-Mail: christoph.jacob@kit.edu},
and Markus Reiher$^{\rm a,}$\footnote{Corresponding Author; E-Mail: markus.reiher@phys.chem.ethz.ch}
}\\

$^{\rm a}$ETH Zurich, Laboratorium f{\"u}r Physikalische Chemie, 
Wolfgang-Pauli-Str.\ 10,\\
CH-8093 Zurich, Switzerland \\
$^{\rm b}$Karlsruhe Institute of Technology (KIT), Center for Functional Nanostructures,\\
Wolfgang-Gaede-Stra\ss{}e 1a, 76131 Karlsruhe, Germany\\[2ex]

\end{center}

\begin{center}
{\large\bf Abstract}\\[1ex]
\end{center}

{\small
The reconstruction of the exchange--correlation potential from accurate \textit{ab initio} electron densities can
provide insights into the limitations of the currently available approximate functionals and provide guidance for 
devising improved approximations for density-functional theory (DFT). For open-shell systems, the spin density
is introduced as an additional fundamental variable in Spin-DFT. Here, we consider the reconstruction of the
corresponding unrestricted Kohn--Sham potentials from accurate \textit{ab initio} spin densities. In particular,
we investigate whether it is possible to reconstruct the spin exchange--correlation potential, which determines
the spin density in spin-unrestricted Kohn--Sham-DFT, despite the numerical difficulties inherent to the optimization of
potentials with finite orbital basis sets. We find that the recently developed scheme for
unambiguously singling out an optimal optimized potential [\textit{J. Chem. Phys.}, \textbf{135}, 244102 (2011)]
can provide such spin potentials accurately. This is demonstrated for two test cases, the lithium atom and the
dioxygen molecule, and target (spin) densities from Full-CI and CASSCF calculations, respectively. }

\vfil

\begin{tabbing}
Date:   \quad \= November 21, 2012  \\
Status:       \> submitted to \textit{J. Chem. Phys.}
\end{tabbing}

\newpage

\section{Introduction}

Density-functional theory (DFT) within the Kohn--Sham (KS) framework \cite{parr,dreizler} represents one of the most 
frequently applied quantum-chemical methods for electronic structure calculations and for the determination
of molecular properties. Its success relies on the accuracy of existing approximations to the exchange--correlation 
energy functional $E_{\rm xc}[\rho]$ and to the exchange--correlation potential $v_{\rm xc}[\rho] = \delta E_{\rm xc}[\rho]/\delta\rho$, 
i.e., the functional derivative of $E_{\rm xc}[\rho]$ with respect to the electron density $\rho(\mathbf{r})$ \cite{koch}. However,
for open-shell systems, in particular for transition metal complexes, the existing approximations have a number of severe
shortcomings \cite{pccp_dft_rev,markus_fd,markus_chimia_2009}, for instance for the prediction of the energy differences 
between different spin states \cite{reiher_reparameterization_2001,reiher_theoretical_2002,harvey_dft_2004,
ye_accurate_2010,swart_spin_2012} and of spin-density distributions  \cite{ghosh2,ghosh,pierloot,feno}.

While the universal functionals $E_{\rm xc}[\rho]$ and $v_{\rm xc}[\rho]$ are unknown, there exists a numerical recipe for obtaining 
the exact exchange--correlation potential $v_{\rm xc}[\rho_0]$ corresponding to the ground-state electron density $\rho_0$ of
arbitrary atomic and molecular systems. First, this ground-state electron density $\rho_0$ can be calculated accurately --- and 
in principle exactly --- using wave-function based \textit{ab initio} calculations. Second, the Kohn--Sham potential 
$v_s[\rho_0]$ that yields the density $\rho_0$ in a noninteracting system can be reconstructed. Finally, by subtracting 
the known nuclear and Coulomb potentials from this reconstructed potential, the exchange--correlation potential 
$v_{\rm xc}[\rho_0]$ can be obtained. Such reconstructed ground-state exchange--correlation potentials can provide guidance 
for the construction of approximate exchange--correlation potentials \cite{baerends94,Gritsenko1999,KT1-KT2,KT3} and 
energy functionals \cite{peach_modeling_2007,teale_accurate_2010,elkind_energy_2012}. 

The key step in the above recipe is the reconstruction of the potential $v_s[\rho_0]$ from the target density $\rho_0$.
This step corresponds to an inverse problem in quantum chemistry \cite{karwowski}, i.e., the  potential is sought which 
generates a given target density in a noninteracting reference system. This potential reconstruction is also essential
for quantum-chemical subsystem and embedding methods (for a review, see Ref.~\cite{gomes_quantum-chemical_2012}), 
in which it can be used to avoid the need for approximating the nonadditive kinetic-energy \cite{roncero2008,roncero2009,
Elliott2009,Elliott2010,huang2011}, or for developing better approximations for this part of the embedding potential
\cite{fux2010,goodpaster2010,jason2011,de_silva_exact_2012}.

The inverse problem of reconstructing the noninteracting local potential yielding a given target density is equivalent to evaluating 
the functional derivative of the noninteracting kinetic-energy functional $T_s[\rho]$ \cite{liu_functional_2004,logdist-2007}, which 
is an implicit functional of the electron density. For evaluating such functional derivatives of implicit functionals, the optimized effective 
potential (OEP) method can be employed, which tackles the inverse problem by minimizing the implicit functional with
respect to the local potential \cite{oep,grling_kohn-sham_1992,WuOEP}, possibly subject to additional constraints \cite{Wu2003}. 
Thus, the reconstruction of local potentials is a special case of the more general problem of evaluating the functional 
derivative of implicit density functionals. 

For applying OEP methods to many-electron systems and large molecules, both the orbitals and the local potential 
have to be expanded in finite basis sets. The introduction of a finite orbital basis set, however, turns the OEP method into 
an ill-posed problem and the solution becomes non-unique \cite{Staroverov2006}. The ill-posed nature is common to many 
inverse problems and makes the inverse mapping from an electron density to a local potential unstable and sensitive to 
optimization parameters. Furthermore, these drawbacks result in unphysical potentials, which can contain large oscillations  
affecting orbital energies and derived properties \cite{Staroverov2006,Heaton-Burgess2007}.

To allow for a routine application of the OEP method in quantum chemistry, approaches to regularize the OEP solutions
have been proposed. One approach developed by He\ss{}elmann $et\,al.$ is based on explicitly constructing an orbital basis 
set that is balanced with respect to the basis set employed for expanding the potential \cite{Hesselmann2007}, whereas
the orbital basis set is balanced implicitly in the approach of Kollmar and Filatov \cite{kollmar_optimized_2007,Fernandez2010}. 
However, these methods require very large orbital basis sets, which hampers their application to larger molecular systems.
A different approach was developed by Yang and co-workers \cite{Bulat2007,Heaton-Burgess2007}, who introduced a regularization 
parameter in the energy functional to make the resulting optimized potentials are as smooth as possible. Recently, an
approach which yields unambiguous potentials for any combination of orbital and potential basis sets and that provides high-quality
potentials already with small orbital basis sets was suggested by one of us \cite{jacob2011}. It is based on the condition that the
optimal reconstructed potential should yield the target density when extending the orbital basis set. 

So far, OEP methods for reconstruction the exchange--correlation potential from accurate \textit{ab initio} densities have 
mainly been applied to closed-shell systems, i.e., to the total electron density as target only (for exceptions, see, 
Refs.~\cite{gritsenko_spin-unrestricted_2004,goodpaster2010}).  For open-shell systems, one commonly employs an
unrestricted KS-DFT formalism \cite{barth,parr,spindft}, in which the spin density $Q(\mathbf{ r}) = \rho^\alpha(\mathbf{ r}) - \rho^\beta(\mathbf{ r})$
is used as an additional fundamental variable. This leads to separate KS equations for $\alpha$- and $\beta$-electrons
containing different exchange--correlation potentials $v_{\rm xc}^\alpha[\rho,Q] = \delta E_{\rm xc}[\rho,Q]/\delta\rho_\alpha$ 
and $v_{\rm xc}^\beta[\rho,Q]= \delta E_{\rm xc}[\rho,Q]/\delta\rho_\beta$, i.e., the $\alpha$- and $\beta$-electron densities 
$\rho^\alpha$ and $\rho^\beta$, respectively, are determined separately. 
In such a formalism, the exact spin-resolved exchange--correlation functional would 
yield --- in addition to the exact total electron density --- also the exact spin density \cite{pople95,perdew09,spindft}. While the total
electron density is determined by the total exchange--correlation potential $v_{\rm xc}^{\rm tot}[\rho,Q] = \frac{1}{2}(v_{\rm xc}^\alpha[\rho,Q] 
+ v_{\rm xc}^\beta[\rho,Q])$, the spin density in unrestricted KS-DFT is determined by the spin exchange--correlation potential 
$v_{\rm xc}^{\rm spin}[\rho,Q] = \frac{1}{2}(v_{\rm xc}^\alpha[\rho,Q] - v_{\rm xc}^\beta[\rho,Q])$ \cite{spindft}. Thus, for improving the spin-density 
dependance of approximate exchange--correlation potentials, it would be desirable to be able to reconstruct this spin exchange--correlation 
potential $v_{\rm xc}^{\rm spin}[\rho_0,Q_0]$ from accurate \textit{ab initio} total and spin densities.

When using finite orbital basis sets, such a reconstruction of $v_{\rm xc}^{\rm spin}[\rho_0,Q_0]$ is particularly challenging,
as it will be more sensitive to numerical errors than for the total or the individual $\alpha$- and $\beta$-electron potentials.
Here, we extent the unambiguous potential reconstruction developed in Ref.~\cite{jacob2011} to the spin-unrestricted cases 
and apply it to the reconstruction of the spin exchange--correlation potential from accurate \textit{ab initio} (spin) densities.
This requires some numerical enhancements of our implementation prompted by the use of Gaussian-type orbitals (GTOs) 
for  expanding the orbitals in such wave-function based calculations. Moreover, the quality of the reconstructed spin potentials 
needs to be assessed carefully, if possible by comparison to reference potentials reconstructed in fully numerical calculations.

This work is organized as follows. Section~\ref{Sec:theory} briefly reviews the potential reconstruction algorithm
and outlines its extension to the spin-unrestricted case. In Section~\ref{Sec:compdet}, the computational methodology 
and extensions of our implementation are described. Subsequently, we study the reconstructed exchange--correlation 
potentials for two test cases, the lithium atom and the O$_2$ molecule, in Section~\ref{Sec:results}. Here, target densities 
from both KS-DFT calculations and from accurate wave-function based \textit{ab initio} calculations (Full-CI and CASSCF) 
are employed. Finally, our conclusions are summarized in Section~\ref{Sec:conclusion}.

\section{Theoretical Background\label{Sec:theory}}

\subsection{Determining Optimized Unrestricted Kohn--Sham Potentials}

Within spin-unrestricted KS-DFT, the wavefunction of the KS reference system is given by a (spin-unrestricted) 
$N$-electron Slater determinant, which is constructed from $N=N^\alpha+N^\beta$ orthonormal one-particle 
functions $\{\phi_i^\sigma(\mathbf{r})\sigma(s), \sigma = \alpha,\beta\}$. The corresponding spatial orbitals $\phi_i^\sigma(\mathbf{r})$
can then by determined by solving two separate sets of one-electron equations \cite{spindft},
\begin{equation}
  \label{eq:ks-eps}
  \left[ -\frac{1}{2}\Delta + v_s^\alpha(\mathbf{r}) \right] \phi^\alpha_i(\mathbf{r}) = \varepsilon^\alpha_i \phi^\alpha_i(\mathbf{r}) 
  \qquad \text{and} \qquad
  \left[ -\frac{1}{2}\Delta + v_s^\beta(\mathbf{r}) \right] \phi^\beta_i(\mathbf{r}) = \varepsilon^\beta_i \phi^\beta_i(\mathbf{r}),
\end{equation}
and the $\alpha$- and $\beta$-electron densities are given by
\begin{equation}\label{edens}
\rho^\alpha(\mathbf{r}) =\sum_i^{N^\alpha} |\phi_i^\alpha(\mathbf{r})|^2 
\qquad
\text{and}
\qquad
\rho^\beta(\mathbf{r}) =\sum_i^{N^\beta} |\phi_i^\beta(\mathbf{r})|^2.
\end{equation}
Here, we consider the inverse problem of determining the spin-resolved KS potential [i.e., the local potentials 
$v_s^\alpha(\mathbf{r})$ and $v_s^\beta(\mathbf{r})$] from given $\alpha$- and $\beta$-electron target 
densities, $\rho^\alpha_0(\mathbf{r})$ and $\rho^\beta_0(\mathbf{r})$, that is, we require
\begin{equation}\label{edens-cond}
\rho^\alpha(\mathbf{r})= \rho^\alpha_0(\mathbf{r}) 
\qquad \text{and} \qquad 
\rho^\beta(\mathbf{r})= \rho^\beta_0(\mathbf{r}).
\end{equation}

An alternative way of expressing this problem is to consider the total and spin densities \cite{spindft},
\begin{equation}
  \rho_0(\mathbf{r}) = \rho^\alpha(\mathbf{r}) + \rho^\beta(\mathbf{r})
  \qquad \text{and} \qquad
  Q(\mathbf{r}) = \rho^\alpha(\mathbf{r}) - \rho^\beta(\mathbf{r})
\end{equation}
as target, and to regard the total and spin potentials \cite{spindft},
\begin{equation}
  v_s^\text{tot}(\mathbf{r}) = \frac{1}{2} \left(  v_s^\alpha(\mathbf{r}) +  v_s^\beta(\mathbf{r})\right)
  \qquad \text{and} \qquad
  v_s^\text{spin}(\mathbf{r}) = \frac{1}{2} \left(  v_s^\alpha(\mathbf{r}) -  v_s^\beta(\mathbf{r})\right) 
\end{equation}
as the quantities that are sought. Here, $v_s^\text{tot}(\mathbf{r})$ is the potential determining the total 
electron density, whereas $v_s^\text{spin}(\mathbf{r})$ determines the spin density. Therefore, this 
representation will be particularly useful for understanding the dependence the exchange--correlation 
potential on the spin density $Q(\boldsymbol{r})$ and to identify the reason for the failure of approximate 
exchange--correlation functionals to describe the spin density correctly in some cases \cite{feno}.   

In principle, any method applicable for reconstructing the KS-potential in closed-shell systems could be adapted to
the spin-unrestricted case by applying it separately to the $\alpha$- and $\beta$-electron densities. However, already the 
closed-shell cases poses many numerical difficulties, and achieving uniform accuracy for the $\alpha$- and $\beta$-spin
potentials, as it is required for obtaining $v_s^\text{spin}(\mathbf{r})$ accurately, turns out to be a challenging task.

The conceptually simplest approach for determining the local potential yielding a given target density is to represent 
the potential numerically on a grid and to determine it iteratively. Several methods working along these lines
have been developed over the past decades \cite{Wang-Parr-1993,baerends94,FAPot,savin-potential,
kadantsev-potential}. 
Generally, these methods calculate the density from some trial potential and then update the potential by comparing the density 
to the target density. If the density is too large at a grid point, the potential is made more repulsive at this point. Conversely, if the 
density is too small, the potential is made more attractive. This process is repeated iteratively until the target density is obtained. 
Different numerical potential reconstruction methods differ in the way in which the potential is updated in each 
iteration \cite{baerends94,savin-potential,kadantsev-potential}. The only exception is the method of Zhao--Morrison--Parr 
(ZMP) \cite{FAPot}, which uses a conceptually different approach.

However, such numerical methods also require that the KS equations [Eq.~\eqref{eq:ks-eps}] are solved numerically on
the same grid. Therefore, their application has mainly been limited to (closed-shell) atoms and, in some cases, (closed-shell) 
diatomic molecules \cite{baerends95,schipper_one_1998}. Here, we will employ such a fully numerical scheme to obtain accurate 
reference potentials for atoms. For determining optimized KS potentials in general molecular systems, both the orbitals and 
the potential are usually expanded in a basis set \cite{grling_kohn-sham_1992,Wu2003}. However, as will be discussed 
below, with finite orbital basis sets the potential reconstruction turns into an ill-posed problem, in which the resulting 
potential is not unique \cite{hirata_can_2001,Staroverov2006,Heaton-Burgess2007}.

To overcome the resulting numerical difficulties, we will apply the recently developed unambiguous optimization 
method \cite{jacob2011} and generalize it to unrestricted KS potentials. This scheme is based on a two-step procedure, 
in which one first determines a non-unique potential using a direct optimization in a finite basis set. Subsequently, an 
unambiguous optimized potential is singled out by means of a suitable criterion.

\subsection{Direct Optimization of Unrestricted Kohn--Sham Potentials}

In the first step, two non-unique local potentials $v^\alpha_s(\mathbf{r})$ and $v^\beta_s(\mathbf{r})$ yielding the target 
$\alpha$- and $\beta$-electron densities $\rho^\alpha_0(\mathbf{r})$ and $\rho^\beta_0(\mathbf{r})$, respectively, in a
given finite basis set have to be determined. To this end, we apply the direct optimization method by Wu and 
Yang \cite{Wu2003} and extend it to the spin-unrestricted case.

The KS kinetic energy for any pair of $\alpha$- and $\beta$-electron densities $\rho^\alpha_0,\rho^\beta_0$
is defined as \cite{parr,oliver}
\begin{align}\label{ks-ekin}
T_s[\rho^\alpha_0,\rho^\beta_0] &= \min_{\Psi_s \rightarrow \rho^\alpha_0,\rho^\beta_0}
\langle \Psi_s | \hat{T} | \Psi_s \rangle  =  \frac{1}{2}T_s[2\rho^\alpha_0] + \frac{1}{2}T_s[2\rho^\beta_0]
\end{align}
where $\hat{T} = - \Delta/2$ is the kinetic-energy operator and where the minimization includes all 
wavefunctions $\Psi_s$ corresponding to a spin-unrestricted $N$-electron Slater 
determinant with $\alpha$- and $\beta$-electron densities $\rho^\alpha(\mathbf{r})$ and $\rho^\beta(\mathbf{r})$. 
Hence, $T_s[\rho^\alpha_0,\rho^\beta_0]$ corresponds to the minimum kinetic energy of an unrestricted KS wave 
function $\Psi_s$ under the constraint that its $\alpha$- and $\beta$-electron densities equal the target 
densities \cite{levy,levy2}.

The constraint minimization problem of Eq.~(\ref{ks-ekin}) can be reduced to two separate
problems for $2\rho^\alpha$ and $2\rho^\beta$, respectively, which results in two Lagrangian
functionals, $W^\alpha[\rho^\alpha(\mathbf{ r})]$ and $W^\beta[\rho^\beta(\mathbf{ r})]$, subject
to the constraints of Eq.~(\ref{edens-cond}) with two corresponding Lagrangian multiplier functions,
$v_s^\alpha(\mathbf{ r})$ and $v_s^\beta(\mathbf{ r})$,
\begin{equation}\label{lfunctional}
W^\sigma[v_s^\sigma] = \sum_i^{N^\sigma} \langle \phi_i^\sigma | \hat{T} | \phi_i^\sigma \rangle
+ \int v_s^\sigma(\mathbf{ r})\big( \rho^\sigma(\mathbf{ r}) - \rho^\sigma_0(\mathbf{ r})  \big) \,{\rm d}^3r \ ~~~{\rm for} ~~ \sigma = \alpha,\beta.
\end{equation}
Following Ref.~\citenum{Wu2003}, the local potentials which yield the target $\alpha$- and $\beta$-electron densities 
can now be determined by the unconstrained maximization of $W^\sigma[\rho^\sigma(\mathbf{ r})]$ with
respect to the local potential $v_s^\sigma(\mathbf{r})$ for each electron spin $\sigma$.

To perform this maximization, the local potential is expanded in a finite basis set as \cite{WuOEP,Wu2003},
\begin{equation}\label{local-pot}
v_s^\sigma(\mathbf{ r}) = v_{\rm ext}(\mathbf{ r}) + v_{\rm Coul}[\rho_0](\mathbf{ r}) + v_0(\mathbf{ r}) + \sum_t b^\sigma_tg_t(\mathbf{ r}),
\end{equation}
where $v_{\rm nuc}(\mathbf{ r})$ is the nuclear potential, $v_{\rm Coul}(\mathbf{ r})$ is the Coulomb potential
of the target density $\rho_0 = \rho_0^\alpha + \rho_0^\beta$, and $v_0(\mathbf{ r})$ represents an initial 
guess for the exchange--correlation potential, while the remainder is expressed as a linear combination 
of a finite set of basis functions $\{g_t(\mathbf{ r})\}$ with coefficients $\{b^\sigma_t\}$.
For fixed $v_{\rm ext}(\mathbf{ r})$ and $v_0(\mathbf{ r})$, the unconstrained maximization of $W^\sigma[v^\sigma]$ turns into an extremum 
problem with respect to the expansion coefficients $\{b^\sigma_t\}$ for each electron spin. The first and second derivatives of $W^\sigma[\rho^\sigma(\mathbf{ r})]$
with respect to $\{b^\sigma_t\}$ can be calculated analytically and one obtains \cite{Wu2003} the following 
expression for the gradient,
\begin{equation}\label{gradient}
\frac{\partial W^\sigma}{\partial b^\sigma_t} 
=  \int g_t(\mathbf{ r})\left( \rho^\sigma(\mathbf{ r}) - \rho^\sigma_0(\mathbf{ r})  \right)  \, {\rm d}^3r 
\end{equation}
and the Hessian, 
\begin{equation}\label{hessian}
H_{st}=\frac{\partial^2 W^\sigma}{\partial b^\sigma_s \partial b^\sigma_t} =  2 \sum_i^{\rm occ^\sigma} \sum_a^{\rm unocc^\sigma} 
\frac{ \langle \phi_i^\sigma | g_s | \phi_a^\sigma \rangle  \langle \phi_a^\sigma | g_t | \phi_i^\sigma \rangle}{\varepsilon_i^\sigma-\varepsilon_a^\sigma},
\end{equation}
for each electron spin $\sigma$. Note that Eqs.~(\ref{gradient}) and (\ref{hessian}) are 
simplified for the case of real-valued orbitals here. With the gradient and Hessian available, the maximization 
can be performed using a standard Newton--Raphson optimization. 


If a finite basis sets is employed for representing the KS orbitals, the potential reconstruction turns into an
ill-posed problem and the optimized potentials resulting from the Wu-Yang direct optimization as described here
are not unique \cite{hirata_can_2001,Staroverov2006,Heaton-Burgess2007}. 
This can be seen \cite{jacob2011} by considering a change in the local $\alpha$- or $\beta$-electron potential 
$\Delta v_s^\sigma(\mathbf{r}) = v_s^\sigma(\mathbf{ r}) - v_{s,0}^{\sigma}(\mathbf{ r})$, 
where $v_{s,0}^\sigma(\mathbf{ r})$ is the potential obtained from the direct optimization, generating the 
orbitals $\{\phi^\sigma_i\}$ and $\{\phi^\sigma_a\}$.  
To first order, this change $\Delta v_s^\sigma(\mathbf{ r})$ induces a response in the density,
\begin{align}\label{deltarho}
 \Delta \rho^\sigma(\mathbf{ r}) 
                      &= 2 \sum_i^{\rm occ^\sigma} \sum_a^{\rm unocc^\sigma} 
                       \frac{\langle \phi^\sigma_a | \Delta v_s^\sigma | \phi^\sigma_i \rangle}{\epsilon^\sigma_i - \epsilon^\sigma_a} \phi^\sigma_i(\mathbf{ r}) \phi^\sigma_a(\mathbf{ r}),
\end{align}
with $\Delta v^\sigma(\mathbf{ r})= \sum_t \Delta b^\sigma_t g_t(\mathbf{ r})$ and $\Delta b^\sigma_t = b^\sigma_t - b_{t,0}^{\sigma}$.
Here, one notices that any change in the potential $\Delta v_s^\sigma(\mathbf{ r})$ will leave the electron density unchanged 
if $\langle \phi^\sigma_a | \Delta v_s^\sigma(\mathbf{ r}) | \phi^\sigma_i \rangle = 0$. Hence, if the orbital basis is not flexible enough, 
the $\alpha$- and $\beta$-electron densities are not affected by certain changes, e.g., oscillations, in the respective potentials. 
Linear combinations of basis
functions $g_t(\mathbf{ r})$ for which the condition $\langle \phi^\sigma_a | \Delta v_s^\sigma(\mathbf{ r}) | \phi^\sigma_i \rangle = 0$ holds
are obtained by inserting the basis set expansion for the potential and performing a singular value decomposition (SVD) of the matrix
$B^\sigma_{ai,t} = \langle \phi^\sigma_a \phi^\sigma_i | g_t \rangle / (\epsilon^\sigma_i - \epsilon^\sigma_a)$, which leads to
\begin{align}\label{deltarho3}
 \Delta \rho^\sigma(\mathbf{ r})
                     &= 2 \sum_r s^\sigma_r \Delta \tilde{b}^\sigma_r \tilde{\Phi}^\sigma_r(\mathbf{ r}),
\end{align}
where $\{s^\sigma_r\}$ are the singular values of $\mathbf{ B}^\sigma$ and where
$\Delta \tilde{b}^\sigma_r = \sum_t V_{t,r}^\sigma \Delta b^\sigma_t$ 
and $\tilde{\Phi}^\sigma_r(\mathbf{r}) = \sum_{ia} U_{ia,r}^\sigma \phi_i^\sigma(\mathbf{r})\phi_a^\sigma(\mathbf{r})$ are the expansion
coefficients of the change in the potential and the occupied--virtual orbital products transformed with the left and right singular vectors,
$(V^\sigma_{t,r})$ and $(U^\sigma_{ia,r})$, respectively. Here, the transformed expansion coefficients $\tilde{b}^\sigma_r$ refer to 
the transformed potential basis functions $\tilde{g}^\sigma_r(\mathbf{ r}) = \sum_t V^\sigma_{t,r} g_t(\mathbf{ r})$. Thus, we notice that 
if one of these transformed potential basis functions $\tilde{g}_r^\sigma(\mathbf{r})$ corresponds to a singular value $s_r$ that is zero or 
very small, the corresponding expansion coefficient $\tilde{b}_r^\sigma(\mathbf{r})$ can be changed (almost) freely without affecting the 
density. Therefore, an additional criterion is necessary for singling out the optimal optimized potential among those yielding the same 
density within the finite basis set.

\subsection{Choosing the Optimal Optimized Potential}
\label{sec:potprojection}

One possibility for unambiguously singling out an optimized potential was suggested in Ref.~\cite{jacob2011}. This 
scheme starts from the requirement that the optimized potential obtained with a finite orbital basis set should be as 
close as possible to the one obtained in the basis set limit. Specifically, the density calculated from the optimal 
optimized potential should still agree with the target density $\rho^\alpha_0(\mathbf{r})$ or $\rho^\beta_0(\mathbf{r})$.
Thus, we introduce a complete set of virtual orbitals (see Ref.~\cite{jacob2011} for details),
\begin{equation}\label{fullorb}
 \tilde{\phi}^\sigma_\mathbf{ r}(\mathbf{ r'}) = \delta(\mathbf{ r} - \mathbf{ r'})- \sum_j^{\rm occ^\sigma} \phi^\sigma_j(\mathbf{ r'})\phi^\sigma_j(\mathbf{ r}),
\end{equation}
with the Dirac delta function $\delta(\mathbf{ r} - \mathbf{ r'})$ and where the second term ensures the orthogonality of 
$\tilde{\phi}^\sigma_\mathbf{r}(\mathbf{r'})$ and the occupied orbitals ${\phi^\sigma_i(\mathbf{ r})}$. With this complete
representation of the virtual orbital space, the change in the electron density due to a variation in the potential [Eq.~\eqref{deltarho}]
can be approximated as
\begin{equation}\label{deltarhofull}
 \Delta \rho^\sigma(\mathbf{ r}) \approx \sum_i^{\rm occ^\sigma} \phi^\sigma_i(\mathbf{ r}) \langle \tilde{\phi}^\sigma_\mathbf{ r} | \hat{T} + v_s^\sigma | \phi^\sigma_i \rangle.
\end{equation}

For singling out the optimal potential, we search for the potential for which the electron density does not change 
considerably when the orbital basis is enlarged, and minimize
\begin{equation}\label{deltarhofullmin}
 \int \frac{\Delta\rho(\boldsymbol{r})^2}{\rho(\boldsymbol{r})} \, {\rm d}^3r
 \approx
 \int \frac{1}{\rho^\sigma(\boldsymbol{r})} \bigg[\sum_i^{\rm occ^\sigma} \phi^\sigma_i(\mathbf{ r}) 
                                                                 \langle {\phi}^\sigma_\mathbf{ r} | \hat{T} + v_s^\sigma | \phi^\sigma_i \rangle \bigg]^2 \, {\rm d}^3r 
 \rightarrow \min.
\end{equation}
Here, the inverse density has been introduced as a weighting function (i.e., the relative change in the density is minimized) 
to obtain a uniformly accurate potential. As is discussed in Ref.~\cite{jacob2011}, this choice can also be justified using theoretical arguments.
The minimization then leads to the linear systems of equations $\mathbf{A}^\sigma \Delta\mathbf{b}^\sigma = - \mathbf{z}^\sigma$ with \cite{jacob2011},
\begin{equation}
  A_{st^\sigma} = \sum_{ij}^{\rm occ^\sigma} \int \frac{\phi_i^\sigma(\mathbf{r})\phi_j^\sigma(\mathbf{r})}{\rho^\sigma(\mathbf{r})} 
                \big\langle \tilde{\phi}^\sigma_{\mathbf{r}} \big| \tilde{g}_s^\sigma \big| \phi_i^\sigma \big\rangle 
                \big\langle \tilde{\phi}^\sigma_{\mathbf{r}} \big| \tilde{g}_t^\sigma \big| \phi_j^\sigma \big\rangle \, {\rm d}^3r
\end{equation}
and
\begin{equation}
 \label{eq:rhs-z}
 z_t^\sigma = \sum_{ij}^{\rm occ^\sigma} \int \frac{\phi_i^\sigma(\mathbf{r})\phi_j^\sigma(\mathbf{r})}{\rho^\sigma(\mathbf{r})} 
                \big\langle \tilde{\phi}^\sigma_{\mathbf{r}} \big| \hat{h}_0^{\sigma} \big| \phi_i^\sigma \big\rangle 
                \big\langle \tilde{\phi}^\sigma_{\mathbf{r}} \big| \tilde{g}_t^\sigma \big| \phi_j^\sigma \big\rangle \, {\rm d}^3r
\end{equation}
where $\hat{h}_0^\sigma = -\Delta/2 + v_s^\sigma(\boldsymbol{r})$. This problem can be solved directly, without explicitly solving the 
KS equations using an extended orbital basis set. Results obtained using this scheme will be referred to as ``\textit{optimal (full)}'' in 
the following.

For comparison, we will also employ two additional schemes for singling out one optimized potential.  The first one, called ``\textit{balanced}''
in the following, is based on the idea that a unique potential is also obtained if the potential basis set is chosen such that it is balanced with
respect to the orbital basis set \cite{Hesselmann2007}. This can be achieved by only retaining those transformed potential basis functions 
$\tilde{g}_r^\sigma(\mathbf{r})$ corresponding to singular values $s_r$ that are not too small, i.e., above a chosen 
threshold $s_\text{thr}$. This is closely related to the OEP scheme of Kollmar and Filatov \cite{kollmar_optimized_2007}. 
Note that in the spin-unrestricted case considered here, such a scheme effectively employs different potential basis sets for the 
$\alpha$- and $\beta$-spin potentials.

In addition, we also use a criterion for singling out the optimized potentials that are as smooth as 
possible (labelled ``\textit{smooth}'' in the following). To this end, we minimize the norm of the gradient of the potential,
\begin{equation}\label{smooth}
 \int \big|\nabla v_s^\sigma(\mathbf{ r}) \big|^2  {\rm d}^3r 
 =  \int \Big[ \sum_t {b}^\sigma_{t} \nabla {g}^\sigma_t(\mathbf{ r}) \Big]^2  {\rm d}^3r \rightarrow \min,
\end{equation}
under the constraint that the change in the density [Eq.~\eqref{deltarho3}] is below a chosen threshold $e_\text{thr}$. This
results in a quadratic programming problem that can be solved using standard approaches \cite{jacob2011}. This criterion is
in close analogy to the method of Yang and coworkers \cite{Heaton-Burgess2007,Bulat2007,Heaton-Burgess2008}, who 
introduced a similar constraint by employing a penalty function during the direct optimization. Note that a common feature of all three
approaches presented here is that they are applied \emph{a posteriori}, and hence a direct optimization of the potentials must be 
performed first. This first step then provides a non-unique potential and corresponding orbitals, which are 
required for the following second step. 

\section{Computational Methodology\label{Sec:compdet}}
All finite basis set calculations were performed with a local version of the Amsterdam Density Functional (\textsc{ADF}) 
program package \cite{adf} together with the \textsc{PyAdf} scripting framework \cite{pyadf}. To allow for the treatment
of spin-unrestricted target densities, we extended our recent implementation \cite{fux2010} of the Wu--Yang
direct optimization algorithm and of the subsequent step for singling out an unambiguous optimized potential \cite{jacob2011}. 

The TZ2P and QZ4P Slater-type orbital (STO) basis sets of \textsc{ADF} were used as orbital basis. 
The potential was expanded in a finite basis set [Eq.~\eqref{local-pot}], using \textsc{ADF}'s density 
fitting basis sets corresponding to the TZ2P or QZ4P orbital basis sets. In all calculations, these basis
sets were augmented with additional $1s$ functions in an even-tempered fashion (see Supporting Information for details). 
As initial guess for the potential, we used a scaled version of the Fermi--Amaldi potential \cite{FAPot,jacob2011} of 
the fixed target electron density $\rho_{0}(\mathbf{ r}) = \rho_{0}^\alpha(\mathbf{ r}) + \rho_{0}^\beta(\mathbf{ r})$, namely
\begin{equation}\label{v0}
v_0(\mathbf{ r}) = v_\text{FA}[\rho_0] (\mathbf{ r}) = - \frac{\xi}{N}  \int \frac{\rho_{0}(\mathbf{ r'})}{|\mathbf{ r}-\mathbf{ r'}|} \, {\rm d}^3r',
\end{equation}
where $\xi$ represents the most diffuse exponent in the STO orbital basis set. This scaled Fermi--Amaldi 
potential ensures that the optimized local potentials have the correct long-range behavior. In the case of
target densities obtained from wave-function based \textit{ab initio} calculations, which employed Gaussian-type orbital (GTO) basis
sets, the Coulomb potential $v_\text{Coul}[\tilde{\rho}_0] $ and Fermi--Amaldi potential $v_\text{FA}[\tilde{\rho}_0] $ 
in Eq.~\eqref{local-pot} are evaluated for an approximate reference density $\tilde{\rho}_0(\mathbf{r})$ obtained 
from a DFT calculation in ADF in which the orbitals are expanded in STO basis functions.

If large basis sets are employed for the potential, the Hessian matrix of Eq.~(\ref{hessian})
contains many small eigenvalues which decay gradually to zero. This causes convergence 
problems during the Newton--Raphson optimization, which we previously addressed by ignoring 
eigenvalues below a certain threshold.
However, for \textit{ab initio} target densities expanded in GTOs, this scheme still caused poor
convergence behavior. Therefore, we followed the work of Wu and Yang \cite{Wu2003b} and
performed a singular value decomposition (SVD) of the Hessian $\mathbf{ H}$. Then the
inverse Hessian can be expressed as
\begin{equation}\label{hessian-inv}
\mathbf{ H}^{-1} = \mathbf{U} \, {\rm  diag}(1/\sigma_r) \, \mathbf{V}^T,
\end{equation}
where the columns of $\mathbf{ U}$ and $\mathbf{ V}$ are the left and right singular vectors, 
respectively, for the corresponding singular values $s_r$. To this inverse Hessian, a Tikhonov 
regularization \cite{tikhonov,NR} is applied by replacing it by
\begin{equation}\label{hessian-inv-f}
\mathbf{ H}^{-1} = \mathbf{U} \, {\rm  diag}(f_r/\sigma_r) \, \mathbf{V}^T,
\end{equation}
where $f_r$ is a filter factor, which is chosen as
\begin{equation}\label{filter}
f_r = \frac{\sigma_r^2}{\sigma_r^2+\lambda^2}.
\end{equation}
We found that appropriate values for the parameter $\lambda$ turn out to be $10^{-4} \leq \lambda \leq 0.01$. 
If $\sigma_r \gg \lambda$, the filter factor $f_r$ is approximately one, while in the case of $\sigma_r \ll \lambda$, $f_r$ approaches 
zero. Thus,  instead of abruptly discarding small singular values, the Tikhonov regularization cuts them off gradually.

As convergence criterion for the Wu--Yang direct optimization, we used the absolute error in the 
$\alpha$- and $\beta$-electron densities $\Delta_{\rm abs}$ compared to the target $\alpha$- and $\beta$-electron
densities, defined as
\begin{equation}\label{deltaabs}
 \Delta^\sigma_{\rm abs} = \int |\rho^{\sigma}(\mathbf{ r}) - \rho^{\sigma}_{0}(\mathbf{ r}) | \,  {\rm d}^3 r.
\end{equation}
The minimal absolute error that can be achieved depends on the considered system, as will be discussed
below.

When singling out the optimal potential according to the scheme of Ref.~\cite{jacob2011}, regions in which
the electron density is very small turn out to be problematic. This is because for densities expanded in
finite GTO or STO basis sets, the exact potential in these regions shows artifacts caused by unphysical
nodes in the density \cite{peach_evaluation_2012,de_silva_pure-state_2012}. To avoid these artifacts, 
grid points  at which the reconstructed $\alpha$- or $\beta$-electron densities are smaller than a threshold 
are  ignored when constructing the right-hand side $\boldsymbol{z}^\sigma$ according to Eq.~\eqref{eq:rhs-z}. 
This corresponds to assuming that the optimized potentials are already well approximated by the initial 
guess if the difference between the target and reconstructed density is smaller than the threshold. For the
lithium atom, this threshold was chosen as $10^{-4}$, whereas for the dioxygen molecule it was set to $10^{-8}$.

For target densities expanded in STOs, the exchange--correlation potential is obtained by adding the 
Fermi--Amaldi potential to the part of the potential expanded in basis functions [cf. Eq.~\eqref{local-pot}].
In the case of target densities obtained in GTOs, the final exchange--correlation potentials for $\alpha$- and 
$\beta$-electrons are obtained as
\begin{equation}
  v_\text{xc}^\sigma(\mathbf{r}) 
  =   \big( v_\text{Coul}[\tilde{\rho}_0] (\mathbf{ r}) - v_\text{Coul}[\rho_\text{WY}] (\mathbf{ r}) \big) 
      + v_\text{FA}[\tilde{\rho}_0] (\mathbf{ r}) + \sum_t b_t^\sigma g_t(\mathbf{r}).
\end{equation}
Here, the first term accounts for the difference between the Coulomb potential used as initial guess 
(evaluated for the density $\tilde{\rho}_0$) and the Coulomb potential corresponding to the target 
density $\rho_0$. The latter is approximated by the density $\rho_\text{WY}$ obtained in a STO expansion 
from the Wu--Yang optimization before singling out an unambiguous potential, as this density represents
to the best available STO representation of the GTO target density. 

To obtain numerical reference potential for atoms, we employed a modified van Leeuwen--Baerends 
algorithm \cite{baerends94} in combination with a numerical solution of the KS equations on a logarithmic 
radial grid \cite{andrae_numerical_1997,eickerling_shell_2008}, as described in
Ref.~\cite{jacob2011}. Here, we used the same initial guess for the potential, and updated the potential
iteratively until the absolute error $\Delta_\text{abs}^\sigma$ compared to the target $\alpha$- and $\beta$-electron 
densities was each below $10^{-4}\,{\rm e\, bohr}^{-3}$.

All CASSCF calculations for obtaining accurate \textit{ab initio} target densities were performed with the 
\textsc{Molpro} program package \cite{molpro} using Dunning's cc-pVTZ basis set for all atoms 
\cite{dunning,dunning2}. For the lithium atom, all electrons are correlated in all orbitals [corresponding 
to a full configuration interaction (Full-CI) treatment], while for the oxygen molecule the electron (spin) density
from a CAS(12,12)SCF calculation was employed. Here, we verified that the resulting densities are converged 
with respect to the dimension of the active space. 

\section{Optimized Potentials from Spin Densities \label{Sec:results}}

\subsection{The Lithium Atom \label{Sec:resultsli}}

\subsubsection*{BP86 target (spin) density expanded in STOs}

\begin{figure}[h]
\caption{Target radial (a) total densities and (b) spin densities for the lithium atom obtained from BP86/QZ4P and
Full-CI/cc-pVTZ calculations. The difference between these two target densities is shown in the lower part.
The insets present the difference between the target total and spin densities and the corresponding density
obtained in the finite orbital basis set from the Wu--Yang optimized potentials (upper insets) and from the optimal 
optimized potentials (lower insets). For the Full-CI/cc-pVTZ results obtained with both the TZ2P and QZ4P orbital 
basis sets in the potential reconstruction are included. The corresponding plots of the $\alpha$- and $\beta$-electron 
densities are given in Fig.~1 in the Supporting Information.}
\label{fig:li-dens}
\vspace{3ex}\hspace*{-1.5cm}
\includegraphics[width=1.15\linewidth]{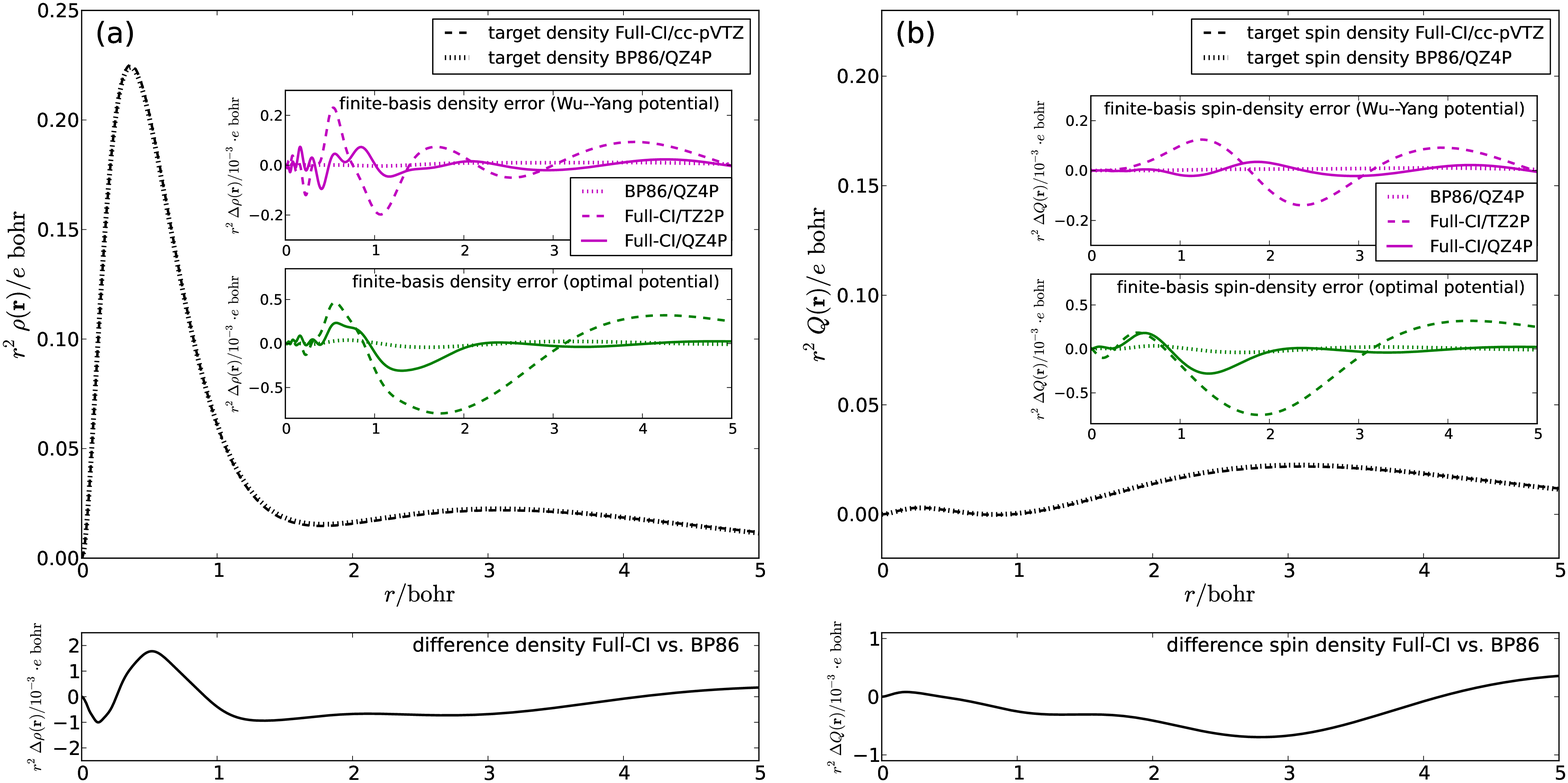}
\end{figure}

As a simple test case, we consider the lithium atom. In its doublet ground-state, there are two $\alpha$-electrons and 
one $\beta$-electron, i.e., one unpaired electron. First, we use the $\alpha$- and $\beta$-electron densities from a 
unrestricted KS-DFT calculation employing the QZ4P orbital basis set and the BP86 exchange--correlation functional 
as target. Here, it should be possible to reconstruct the target (spin) density accurately in a potential
reconstruction using the same orbital basis set. The target total and spin densities are shown in 
Fig.~\ref{fig:li-dens}. For the lithium atom, there is only minimal spin polarization and the spin density is determined
by the unpaired electron in the $2s$ orbital, whereas almost identical $1s$ orbitals are obtained for $\alpha$- and
$\beta$-electrons (see also Fig.~1 in the Supporting Information). 

\begin{figure}[h]
\caption{
Reconstructed potentials determined for the Li atom and a BP86/QZ4P target (spin) density. The upper part shows the 
exchange--correlation potentials for (a) $\alpha$ electrons $v_\text{xc}^\alpha$ and (b) $\beta$ electrons 
$v_\text{xc}^\beta$, while the lower part shows (c) the total exchange--correlation potential $v_\text{xc}^\text{tot}$ 
and (d) the spin exchange--correlation potential $v_\text{xc}^\text{spin}$. 
For the potential reconstruction with the finite QZ4P orbital basis set, the potentials obtained with the different
schemes for singling out an unambiguous potential (see text for details) are shown.
The accurate potentials obtained with a numerical solution of the KS equations (\textit{``numerical (STO)''}) as well
as the BP86 exchange--correlation potential calculated from the reference density (\textit{``BP86 xc potential''}) are 
shown for comparison. The latter is shifted such that it agrees with the numerical reference at $r=5$~bohr.}
\label{fig:lipotentials-bp}
\vspace{3ex}\hspace*{-1.5cm}
\includegraphics[width=1.15\linewidth]{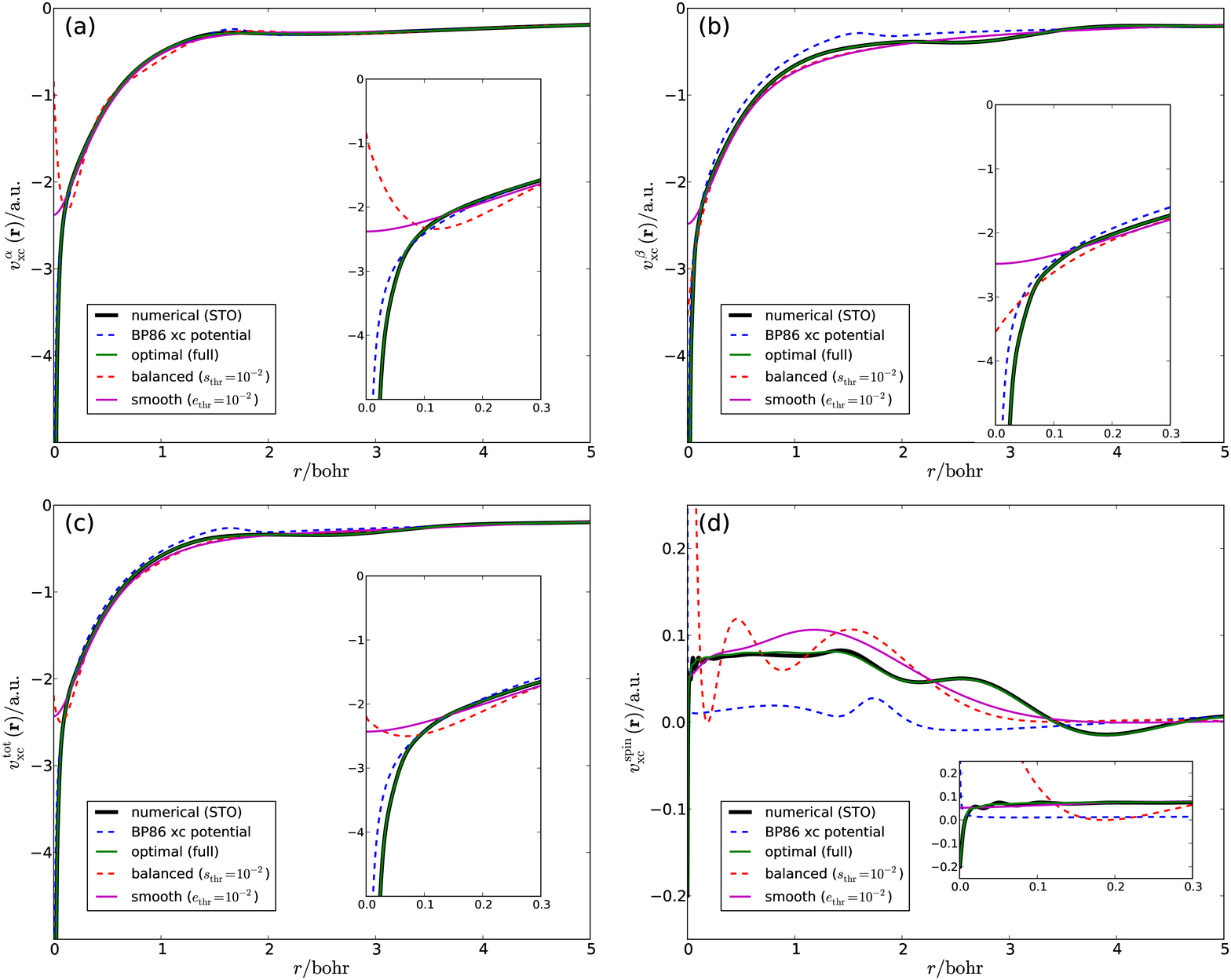}
\end{figure}

For assessing the quality of the optimized potentials obtained with finite orbital basis sets, we determined
the exchange--correlation potentials for $\alpha$- and $\beta$-electrons numerically as described 
above. These are presented in Fig.~\ref{fig:lipotentials-bp} as reference (black line) alongside the potentials obtained from 
potential reconstruction calculations employing the finite QZ4P orbital basis set and the corresponding QZ4P density 
fitting basis set augmented with additional tight $1s$ functions for expanding the potential. In addition, we included the
BP86 exchange--correlation potential evaluated for the target density (blue dashed line), i.e., the potential that
was used in the finite-basis set KS-DFT calculation for determining the target $\alpha$- and $\beta$-electron densities. 
We note that, even though it is close to it, this BP86 potential does not agree with the numerical reference potential. As 
was pointed out before, these two potentials should only be equal in the basis set limit \cite{jacob2011,de_silva_pure-state_2012}.

The reconstructed $\alpha$- and $\beta$-electron exchange--correlation potentials $v_\text{xc}^\alpha(\mathbf{r})$ and
$v_\text{xc}^\beta(\mathbf{r})$ are shown in Figs.~\ref{fig:lipotentials-bp}a and~\ref{fig:lipotentials-bp}b. In both cases,
the potentials obtained from the Wu--Yang optimization in the first step show large oscillations and are,
therefore, not shown in the figures. These oscillations are removed if an unambiguous potential is singled
out in the second step. Irrespective of which of the schemes described in Section~\ref{sec:potprojection} is 
applied, the potentials closely agree with the numerical reference for $r>0.5$~bohr. However, differences 
are found closer to the nucleus (see also the insets in Fig.~\ref{fig:lipotentials-bp}).  When using an implicitly
balanced potential basis set (dashed red line), the potential is too large close to the nucleus. The $\alpha$-electron 
potential has a spurious minimum in this case and show slight oscillations around the numerical reference potential.
Singling out the potential that is as smooth as possible (solid magenta line) does not introduce oscillations, but
also results in a potential with the wrong behavior for small $r$.
On the other hand, the optimal potential determined using the criterion of Eq.~\eqref{deltarhofullmin} closely matches 
the numerical reference potential also close to the nucleus.

\begin{table}[h]
\caption{
Absolute errors $\Delta^\sigma_{\rm abs}$ in the $\alpha$- and $\beta$-electron densities with respect to the 
target $\alpha$- and $\beta$-electron densities (in ${\rm e\, bohr}^{-3}$) obtained with different reconstructed potentials for 
the Li atom with BP86/QZ4P target densities. The QZ4P orbital basis set is used in the finite-basis set potential reconstruction.
$\Delta_{\rm abs}^{\sigma, \rm finite}$ refers to the error in the density obtained from the respective potentials in the 
finite orbital basis set, whereas $\Delta_{\rm abs}^{\sigma, \rm num}$ is the error for the density obtained from a 
numerical solution of the KS equations.}
\label{tab:error_li}
\begin{center}
{\small 
\begin{tabular}{lccccc}\hline \hline
     &  $\Delta_{\rm abs}^{\alpha,\rm num}$ & $\Delta_{\rm abs}^{\alpha,\rm finite}$   & & 
         $\Delta_{\rm abs}^{\beta,\rm num}$ & $\Delta_{\rm abs}^{\beta,\rm finite}$  \\ 
\hline
Numerical (STO)		                   &  $<10^{-4}$ & 0.0007  & 	&  $<10^{-4}$ & 0.0001   \\
Wu--Yang            		                   &  0.0098 & 0.0008 & 		&  0.0036 & $<10^{-4}$  \\
Balanced ($s_{\rm thr}=10^{-2}$)	&  0.0090 & 0.0024 &  		&  0.0016 & 0.0006  \\
Smooth ($e_{\rm thr} = 10^{-2}$)	&  0.0087 & 0.0086 & 		&  0.0082 & 0.0082  \\
Optimal (full)                             		&  0.0009 & 0.0014 & 		& $<10^{-4}$ &  $<10^{-4}$  \\
\hline\hline
\end{tabular}}
\end{center}
\end{table}

For a more quantitative comparison of the different approaches, we present the absolute errors $\Delta_\text{abs}^{\sigma,\text{num}}$
[cf. Eq.\eqref{deltaabs}] in the $\alpha$- and $\beta$-electron densities obtained from the different potentials in a numerical 
solution of the KS equations compared to the target $\alpha$- and $\beta$-electron densities in Table~\ref{tab:error_li}. Naturally, this absolute
error is the smallest for the numerical reference potential. For the $\alpha$-electron potentials, the largest error is obtained for the 
potential obtained from the Wu--Yang optimization with ca.~$10^{-2}$~${\rm e\, bohr}^{-3}$, i.e., the error is two magnitudes larger than 
for the numerical reference. This error is only slightly reduced by using an implicitly balanced basis set or by singling out a smooth 
potential. In contrast, for the optimal potential, the absolute error in the numerical density is reduced by an order of magnitude 
below $10^{-3}$~${\rm e\, bohr}^{-3}$. A similar picture is obtained for the $\beta$-electron potentials, even though all the errors are 
smaller in this case. Thus, these results confirm the previous finding that high-quality potentials can be obtained by applying the
criterion of Eq.~\eqref{deltarhofullmin} for unambiguously singling out the optimal potential \cite{jacob2011}.

In addition to the errors in the densities obtained from a numerical solution of the KS equations with the different potentials,
Table~\ref{tab:error_li} also includes the absolute errors $\Delta_\text{abs}^{\sigma,\text{finite}}$ obtained with these potentials in 
the finite orbitals basis set. In this case, the smallest error is obtained for the potentials obtained directly from the Wu-Yang 
optimization, and these absolute errors correspond to the convergence criterion used in this step. After singling out one
optimized potential in the second step the error increases, but the smallest one is obtained for the optimal potential. Note
that for both the optimal and the numerical reference potential the absolute density errors are larger in the finite orbital
basis set than for the numerical solution of the KS equations. This discrepancy was discussed previously \cite{jacob2011} and 
arises because it is in general not possible to reproduce the target density both in a given finite basis set and
in a fully numerical calculation at the same time. However, as was noted before, these differences decrease
when increasing the size of the orbital basis set. 

After assessing the quality of the reconstructed $\alpha$- and $\beta$-electron exchange--correlation potentials, we 
turn to Figs.~\ref{fig:lipotentials-bp}c and Figs.~\ref{fig:lipotentials-bp}d, where the same results are shown as total 
exchange--correlation potentials 
$v_\text{xc}^\text{tot}(\mathbf{r}) = \frac{1}{2} (v_\text{xc}^\alpha(\mathbf{r})+v_\text{xc}^\beta(\mathbf{r}))$ and spin 
exchange--correlation potentials 
$v_\text{xc}^\text{spin}(\mathbf{r})= \frac{1}{2} (v_\text{xc}^\alpha(\mathbf{r})-v_\text{xc}^\beta(\mathbf{r}))$, respectively.
While for the total potential, the overall results are similar to those discussed for the individual spin components, it is
apparent that reconstructing the spin potential accurately is significantly more difficult. To calculate the difference
between the $\alpha$-electron potential and the $\beta$-electron potential reliably, it is necessary to determine each 
of these individual $\alpha$- and $\beta$-electron potentials with comparable accuracy. 

For the lithium atom considered here, the spin polarization is very small. Therefore, the spin part of the 
exchange--correlation potential is very small as well. In particular, it is almost constant in the range probed by the
$1s$ orbital and, therefore, does not introduce a significant spin polarization for the $1s$ orbital. Thus, the
spin potential is mostly due to the different asymptotic decay of the $\alpha$- and $\beta$-electron densities.
Since the BP86 exchange--correlation potentials evaluated from the target $\alpha$- and $\beta$-electron densities 
do not show the correct asymptotic decay, the corresponding spin potential almost vanishes. 
Even though the spin potential is evaluated as the difference of the much larger $\alpha$- and $\beta$-electron 
potentials, the optimal potential reproduces the numerical reference almost perfectly. 
On the other hand, the smooth and the balanced potential deviate from the numerical reference not only close 
to the nuclei (where these differences were also recognizable for the $\alpha$- and $\beta$-spin potentials), but 
also further away from the nuclei. Thus, in order to reconstruct the spin potential $v_\text{xc}^\text{spin}(\boldsymbol{r})$ 
reliably, it is essential to single out the optimal potential according to the criterion of Eq.~\eqref{deltarhofullmin}.

\subsubsection*{Full-CI target (spin) density expanded in GTOs}

Next, we use the accurate $\alpha$- and $\beta$-electron densities from a Full-CI calculation for the lithium
atom as our target. In this case, the target $\alpha$- and $\beta$-electron densities have to be expanded in a GTO orbital 
basis set, which might result in additional difficulties when performing the potential reconstruction with our implementation
using an STO orbital basis set. The Full-CI total and spin densities are included in Fig.~\ref{fig:li-dens} and are 
on the scale of the plots almost indistinguishable from the BP86 ones considered above. Nevertheless,
the difference densities also included in Fig.~\ref{fig:li-dens} show that there are slight differences, in particular
for the $1s$ orbital density and the asymptotic decay.

\begin{figure}[h]
\caption{Reconstructed potentials determined for the Li atom and a Full-CI/cc-pVTZ target (spin) density with a TZ2P and with a
QZ4P orbital basis set in the potential reconstruction. The upper part shows the (a) total exchange--correlation 
potentials $v_\text{xc}^\text{tot}$ and (b) the spin exchange--correlation potentials $v_\text{xc}^\text{spin}$
obtained with the TZ2P orbital basis set in the potential reconstruction, whereas the lower part (c) and (d)
shows the corresponding results for the QZ4P orbital basis set. 
The accurate potentials obtained with a numerical solution of the KS equations (\textit{``numerical (GTO)''}) from
the GTO target density as well as from the Wu--Yang reconstructed density  (\textit{``numerical (STO)''}) are 
shown for comparison. These reference potentials have been shifted such that they agree with the optimal
potential at $r=2.7$~bohr. The corresponding plots of the $\alpha$- and $\beta$-electron potentials are given
in Fig.~2 in the Supporting Information.}
\label{fig:lipotentials}
\vspace{3ex}\hspace*{-1.5cm}
\includegraphics[width=1.15\linewidth]{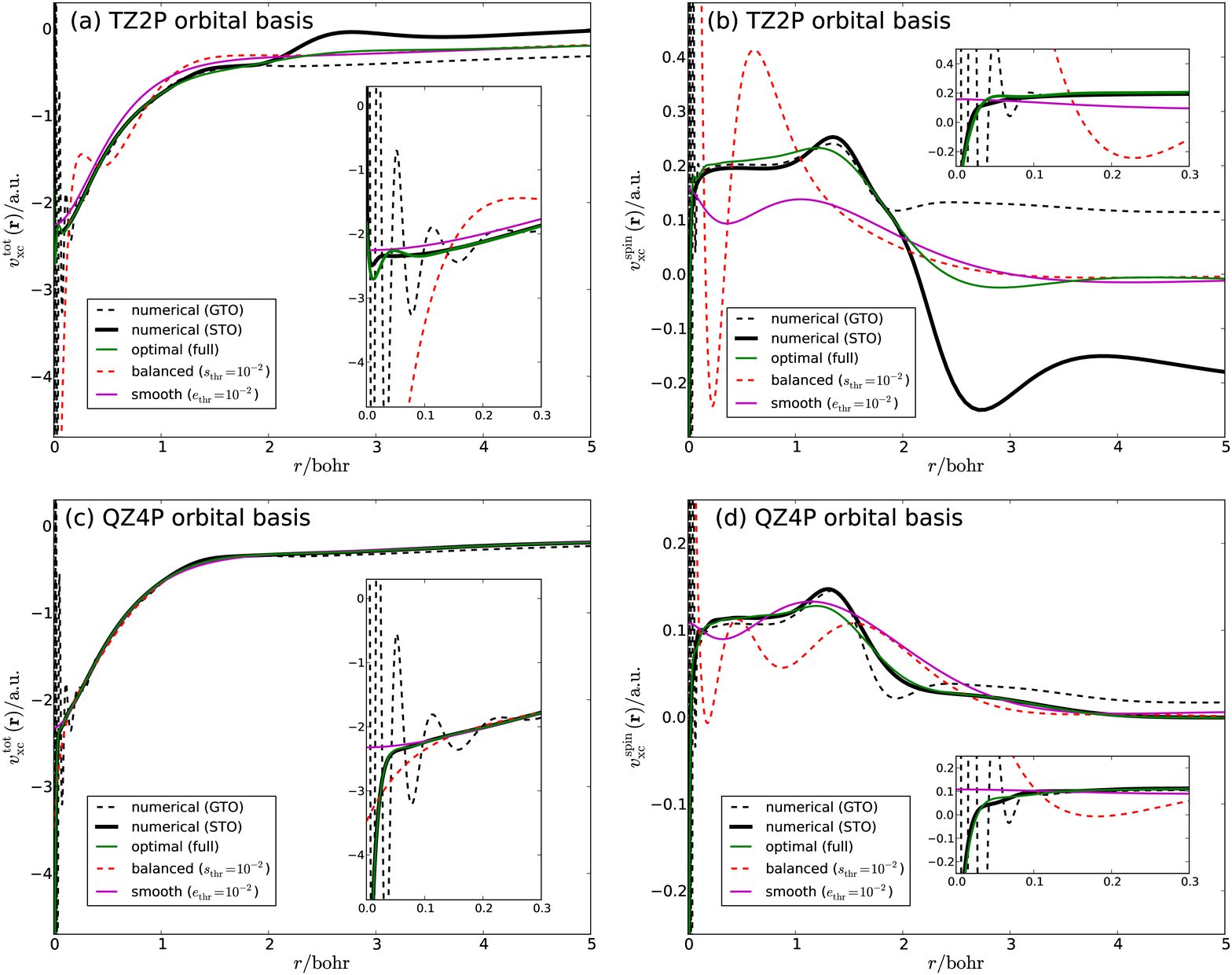}
\end{figure}

Fig.~\ref{fig:lipotentials} shows the reconstructed total and spin exchange--correlation potentials, whereas the 
individual $\alpha$- and $\beta$-electron potentials are presented in Fig.~2 in the Supporting Information.
The potentials obtained from the Full-CI density with a numerical solution of the KS equations
are included as black dashed lines in these plots. These reference potentials features considerable oscillations 
which are most pronounced near the nucleus. Such an oscillatory behavior in the reconstructed potentials is commonly 
found when the target density is expanded in a set of Gaussian-type orbitals (GTOs) \cite{baerends97} and can
be attributed to the deficiency of GTOs to represent the correct form of the electron density close to
the nucleus. However, these oscillations can be reduced when the GTO orbital basis set used 
for determining the target density is enlarged \cite{baerends97}. This is shown in Fig.~3 in the Supporting Information
by comparing numerical potentials reconstructed from the Full-CI/cc-pVTZ target density and from a CAS(3,30)SCF/cc-pVQZ
target density.

As a first step of the potential reconstruction, we determine a (non-unique) potential using the Wu--Yang direct optimization
algorithm. Here, we apply two different STO orbital basis sets (TZ2P and QZ4P) in combination with the corresponding
density fitting basis sets augmented with additional tight $1s$ functions for expanding the potential. While for the BP86
target density expanded in STOs this optimization could be converged until the absolute error in the density was smaller
than $8.0\cdot10^{-4}$ and $10^{-4}$~${\rm e\, bohr}^{-3}$ for the $\alpha$- and $\beta$-electron density, respectively, looser
convergence criteria have to be used for the GTO target densities. With a TZ2P orbital basis set, the absolute density error
in the Wu--Yang optimization only reaches $5.1\cdot10^{-3}$ and $2.0\cdot10^{-3}$~${\rm e\, bohr}^{-3}$ for the $\alpha$- 
and $\beta$-electron density, respectively (see Tab.~\ref{tab:error_li-2}). This is further reduced to  $2.5\cdot10^{-3}$ and 
$4.0\cdot10^{-4}$~${\rm e\, bohr}^{-3}$, respectively, with the larger QZ4P orbital basis set. The corresponding 
difference densities are shown in the upper insets in Fig.~\ref{fig:li-dens}. Thus, a sufficiently large STO orbital basis
set is required to be able to represent the target density from an \textit{ab initio} calculation using GTOs. Nevertheless,
with the QZ4P basis set it is possible to achieve an agreement close to the one obtained for the target density expanded
in STOs. We note that the methodological improvements discussed in Section~\ref{Sec:compdet}, in particular the use of 
the Tikhonov regularization, are essential here to make this convergence possible.

\begin{table}[h]
\caption{
Absolute errors $\Delta^\sigma_{\rm abs}$ in the $\alpha$- and $\beta$-electron densities with respect to the 
target $\alpha$- and $\beta$-electron densities (in ${\rm e\, bohr}^{-3}$) obtained with different reconstructed 
potentials for the Li atom and Full-CI/cc-pVTZ target densities. 
Results obtained both with the TZ2P and with the QZ4P orbital basis set in the potential reconstruction
are shown. $\Delta_{\rm abs}^{\sigma, \rm finite}$ refers to the error in the density obtained from the respective
potentials in the finite orbital basis set, whereas $\Delta_{\rm abs}^{\sigma, \rm num}$ is the error for the
density obtained from a numerical solution of the KS equations.}
\label{tab:error_li-2}
{\small \vspace{4ex}
\hspace{-0.7cm}
\begin{tabular}{lccccc c ccccc}\hline \hline
     &  \multicolumn{5}{c}{TZ2P} && \multicolumn{5}{c}{QZ4P} \\
     \cline{2-6}\cline{8-12}
     Full-CI/cc-pVTZ
     & $\Delta_{\rm abs}^{\alpha,\rm num}$ & $\Delta_{\rm abs}^{\alpha,\rm finite}$   & & 
     $\Delta_{\rm abs}^{\beta,\rm num}$ &     $\Delta_{\rm abs}^{\beta,\rm finite}$   & &
     $\Delta_{\rm abs}^{\alpha,\rm num}$ &     $\Delta_{\rm abs}^{\alpha,\rm finite}$   & & 
     $\Delta_{\rm abs}^{\beta,\rm num}$ &   $\Delta_{\rm abs}^{\beta,\rm finite}$    \\ 
\hline
Numerical (GTO)                        		& $<10^{-4}$ & 0.0258  & 	& $<10^{-4}$ & 0.0086 &
		                                 		& $<10^{-4}$ & 0.0065 & 	& $<10^{-4}$ & 0.0010 \\
Numerical (STO)                         		& 0.0051  & 0.0158 & 	& 0.0020   & 0.0014 &
		                                 		&  0.0025 & 0.0051  & 	&  0.0005 &  0.0005  \\
Wu--Yang                                   		&  0.0782 & 0.0051   & 	&  0.0301 & 0.0020  &
		                                 		&  0.0165 & 0.0025 & 	&  0.0021 & 0.0004  \\
Balanced ($s_{\rm thr}=10^{-2}$)	&  0.0336 & 0.0079  & 	&  0.0310 & 0.0020  &
							&  0.0092 & 0.0074 & 	&  0.0030 & 0.0004  \\
Smooth ($e_{\rm thr} = 10^{-2}$)	&  0.0416 & 0.0105  & 	&  0.0039 & 0.0041  &
							&  0.0120 & 0.0071 & 	&  0.0062  & 0.0061 \\
Optimal (full)                             		&  0.0220 & 0.0284  & 	&  0.0021  & 0.0016 & 
			                      		&  0.0061 & 0.0067 & 	&  0.0005  & 0.0006 \\
\hline\hline
\end{tabular}}
\end{table}

Since the numerical potential reconstructed from the GTO densities show oscillations due to the insufficiencies
of the GTO basis set close to the nucleus, we also performed a numerical potential reconstruction using the $\alpha$-
and $\beta$-electron densities from the Wu--Yang optimization, which are expanded in an STO orbital basis, as target. 
These are included in Fig.~\ref{fig:lipotentials} as solid black line. They do not show oscillations near the nucleus
anymore, but otherwise closely match the numerical potentials reconstructed from the GTO densities. However, 
for the TZ2P orbital basis set, the total and spin potentials reconstructed from the STO and GTO densities differ 
significantly for $r>2$~bohr.
This can be traced back to the $\beta$-electron potential (see Fig.~2 in the Supporting Information),
and is caused by spurious nodes appearing in the $\beta$-electron density in the region where it is very 
small \cite{peach_evaluation_2012,de_silva_pure-state_2012}. The $\beta$-electron density in the lithium atom
is particularly prone to such artifacts because it is due to only one orbital, i.e., nodes in this orbital are not 
offset by other (nonzero) orbitals.  In the following, we will consider the numerical potentials reconstructed
from the Wu--Yang densities expanded in STOs as reference for the finite-basis set potential reconstruction. 

The optimized potentials obtained from the finite-basis set reconstruction using the different schemes for
singling out one unambiguous potential after the Wu--Yang optimization are included in Fig.~\ref{fig:lipotentials}.
The potentials obtained directly from the Wu--Yang optimization contain considerable oscillations and are hence 
not shown in the figure.
First, we consider the total exchange--correlation potentials in Figs.~\ref{fig:lipotentials}a and~\ref{fig:lipotentials}c
(for the individual $\alpha$- and $\beta$-electron potentials, see Fig.~2 in the Supporting Information).
With the TZ2P orbital basis set, both the smooth and the optimal potential show a good agreement with the
numerical reference potential for $r<2$~bohr, whereas the potential obtained with an implicitly balanced 
potential basis set still features some small oscillations. For $r<2$~bohr, none of the potentials can
reproduce the bump in the reference potential caused by artifacts of the $\beta$-electron density in this
region. In principle, the optimal potential should account for this feature, since the criterion of Eq.~\eqref{deltarhofullmin}
minimizes the deviation from the Wu--Yang density in the basis-set limit, i.e., it should converge to the numerical
reference potential determined from that density.
However, as discussed in Section~\ref{Sec:compdet}, regions where the density is smaller than $10^{-4}$~${\rm e\, bohr}^{-3}$,
which is reached for the $\beta$-electron density at $r>2$~bohr, are discarded in our implementation to avoid such 
artifacts at small densities. In fact, we verified that after increasing the threshold for discarding small densities, the 
optimal potential does account for the bump in the reference potential.

When increasing the size of the orbital basis set to QZ4P, the total exchange--correlation potentials from all
three schemes closely agree with the numerical reference. Differences only remain for the smooth and balanced
potentials close to the nucleus, while the optimal potential matches the reference also in this region. For a
more quantitative comparison, the absolute errors $\Delta_{\rm abs}^{\sigma,\rm num}$ in the $\alpha$- and $\beta$-electron 
densities obtained from the different potentials in a numerical solution of the KS equations compared to the target 
$\alpha$- and $\beta$-electron densities are listed in Table~\ref{tab:error_li-2}. In all cases, the smallest error is achieved for the
optimal potential. Moreover, the absolute errors in the numerical densities decrease for all three schemes when 
going from the TZ2P to the QZ4P orbital basis set.

In addition, Table~\ref{tab:error_li-2} includes the absolute errors $\Delta_{\rm abs}^{\sigma,\rm finite}$ in the
$\alpha$- and $\beta$-electron densities obtained with the different potential in the finite orbital basis set. In
general, these errors increase compared to the potentials obtained from the Wu--Yang procedure when applying 
the schemes for singling out one optimized potential. For the TZ2P orbital basis, the largest error in the $\alpha$-electron
density within the finite basis set is in fact obtained for the optimal potential, even though for this potential the
error in the numerical density is the smallest. Again, this is because the error in the finite-basis set density and in the numerical
density cannot be minimized at the same time \cite{jacob2011}. Note that the finite-basis set density error with the
optimal potential is comparable to the one obtained with the numerical reference potential.
When going to the larger QZ4P orbital basis set, the finite-basis set density errors decrease significantly, both for the 
optimal and for the numerical reference potentials. This can also be seen in the finite-basis set difference densities
obtained with the optimal potentials, which are included in the lower insets in Fig.~\ref{fig:lipotentials}.

Finally, we turn to the reconstructed spin exchange--correlation potentials shown in Fig.~\ref{fig:lipotentials}b 
and~\ref{fig:lipotentials}d. Again, reproducing the numerical reference potential is much more difficult in this
case because the spin potential is calculated as the difference of the $\alpha$- and $\beta$-electron potentials.
Therefore, even though with the TZ2P orbital basis set, the smooth potential visually agrees with the reference
for the total and the individual $\alpha$- and $\beta$-electron potentials at $r<2$~bohr (see Fig.~\ref{fig:lipotentials}a
and Figs.~2a and~2b in the Supporting Information), the corresponding spin potential deviates significantly. 
On the other hand, the optimal potential shows a good agreement with the numerical reference for $r<2$~bohr
and only fails to reproduce the spurious behavior for larger $r$ (see discussion above). With the QZ4P orbital
basis set, the agreement with the numerical reference improves for all reconstructed potentials. However, for
the potential obtained with an implicitly balanced basis set, there are still significant oscillations, which are
not obvious in the plots of the individual $\alpha$- and $\beta$-electron potentials (see Figs.~2c and~2d in 
the Supporting Information). The smooth potential qualitatively reproduces all features of the numerical reference,
but the best agreement is achieved for the optimal potential.  

The comparison of the TZ2P and the QZ4P orbital basis sets demonstrates that the optimal potentials obtained 
with the smaller TZ2P orbital basis already agree very well with the numerical reference potential determined from the 
Wu--Yang density, which corresponds to the closest available approximation of the target density in the STO orbital 
basis. However, this Wu--Yang density differs from the target density expanded in GTOs so that deviations to the
numerical potential determined from that target density occur. Thus, the QZ4P orbital basis set is required in order
to obtain a (spin) potential that agrees with the numerical potential calculated from the GTO target density not 
because of the basis set requirements of the scheme for singling out the optimal potential, but because of the need
to reproduce the target density represented in an GTO basis set with STOs in the Wu--Yang optimization step.
However, the use of an STO representation of the density in the first step before determining the optimal potential
according to the criterion of Eq.~\eqref{deltarhofullmin} has the advantage of avoiding the spurious oscillations in the 
reconstructed potential arising for a GTO expansion of the target density close to the nucleus. Of course, a sufficiently
large STO basis set will be able to reproduce the spurious behavior of the GTO density close to the nucleus and
will result in a potential showing the corresponding oscillatory behavior.

Finally, we note that the optimal potentials presented here are converged with respect to the size of the potential
basis set. Adding additional tight or diffuse functions does not alter the resulting optimal potentials significantly
anymore. The dependence of the optimal potentials on the potential basis set is illustrated in 
Figs.~4 and~5 and Table~I and~II in the Supporting Information.

\subsection{An Open-Shell Molecule: Dioxygen\label{Sec:resultso2}}

\subsubsection*{BP86 target (spin) density expanded in STOs}

\begin{figure}[h]
\caption{Target spin densities for the dioxygen molecule obtained from BP86/QZ4P and CAS(12,12)SCF/cc-pVTZ 
calculations. Both spin densities are compared (a) along the bond axis ($x$-axis), (b) perpendicular to the bond 
axis ($y$-axis). Furthermore, (c) and (d) show the BP86/QZ4P and CAS(12,12)SCF/cc-pVTZ spin densities, respectively, 
in the $xy$-plane.
The corresponding plots of the total and the individual $\alpha$- and $\beta$-electron densities are given in 
Fig.~6 in the Supporting Information.}
\label{fig:o2-dens}
\vspace{3ex}\hspace*{-1.5cm}
\includegraphics[width=1.15\linewidth]{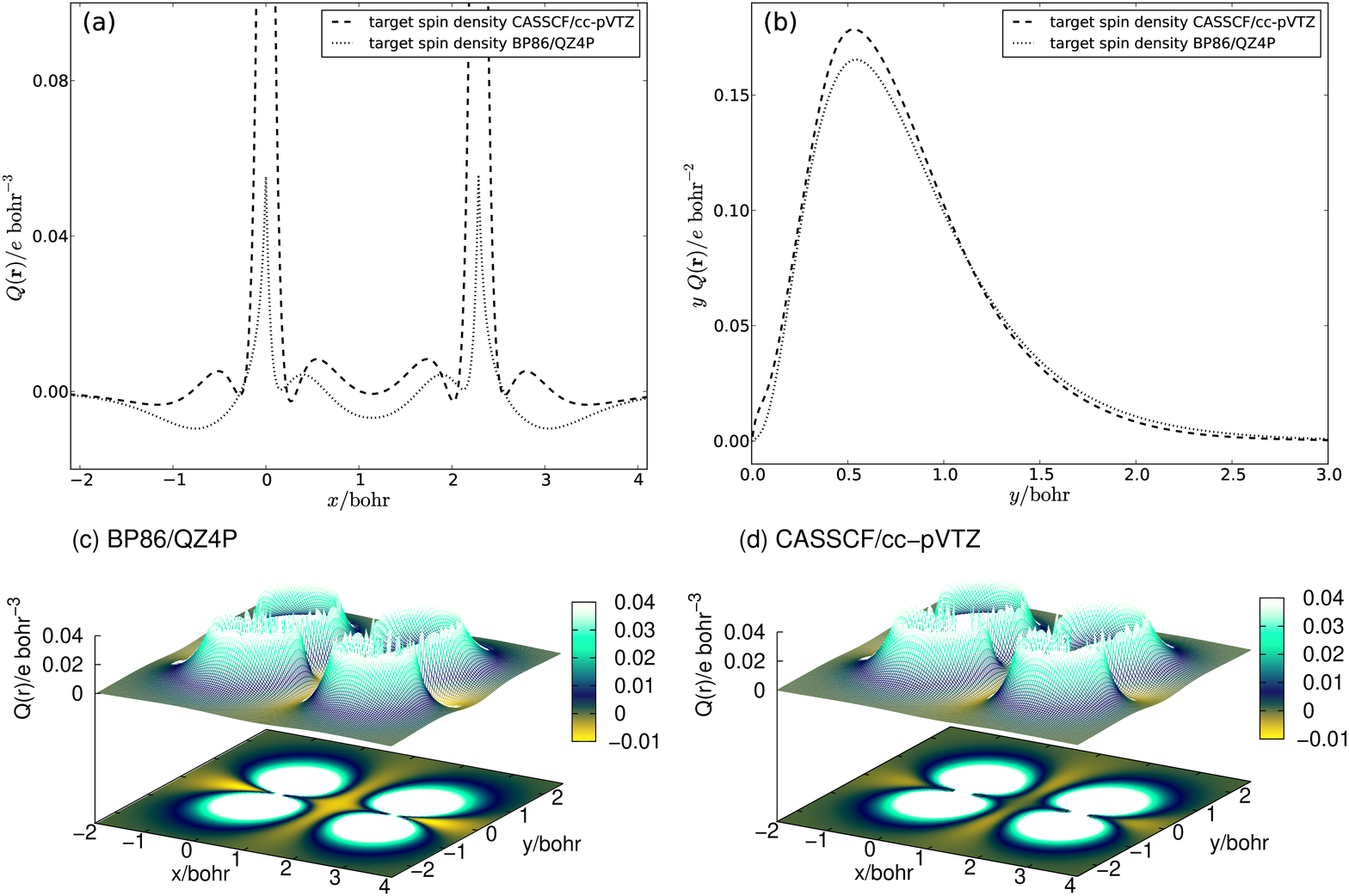}
\end{figure}

We now investigate a diatomic molecule with an open-shell ground state, namely dioxygen O$_2$ with an O--O bond 
distance of 1.21~{\AA} in its equilibrium structure. Here, the two antibonding $\pi^*$-orbitals are singly occupied. 
First, we consider a target density from an unrestricted KS-DFT calculation using the BP86 exchange--correlation functional and 
the Slater-type QZ4P orbital basis set. This target spin density is shown in Fig.~\ref{fig:o2-dens}a and~\ref{fig:o2-dens}b
along the bonding axis ($x$-axis) and perpendicular to it ($y$-axis), respectively. Along the $y$-axis, the spin density
is determined by the singly occupied orbitals which have a cylindrical shape around the bond axis, as is also visible
in the plot of the spin density in the $xy$-plane in Fig.~\ref{fig:o2-dens}c. On the other hand, the singly occupied
orbitals vanish on the bond axis and, therefore, the spin density along the $x$-axis is solely due to the spin polarization
of the doubly occupied orbitals (i.e., the differences between the occupied $\alpha$- and $\beta$-electron orbitals). In particular,
there is a region where the spin density becomes negative (see the yellow regions in Fig.~\ref{fig:o2-dens}c).

\begin{figure}[h]
\caption{Reconstructed potentials determined for the dioxygen molecule and a BP86/QZ4P target (spin) density. The upper part shows 
the total exchange--correlation potential $v_\text{xc}^\text{tot}$ (a) along the bond axis ($x$ axis) and (b) perpendicular
to the bond axis along the $y$ axis. The lower part shows the spin exchange--correlation potential $v_\text{xc}^\text{spin}$ 
along the (c) $x$ axis and (d) $y$ axis. In the potential reconstruction, the finite QZ4P orbital basis set was employed.
For comparison, the BP86 exchange--correlation potential calculated from the reference density (\textit{``BP86 xc potential''}) 
is also included. This BP86 potential is shifted such that it agrees with the optimal potential at $x=-2$~bohr or at $y=+2$~bohr,
respectively. 
The corresponding plots of the individual $\alpha$- and $\beta$-electron potentials are shown in Fig.~7 in the Supporting
Information.}
\label{fig:o2potentials-bp}
\vspace{3ex}\hspace*{-1cm}
\includegraphics[width=1.15\linewidth]{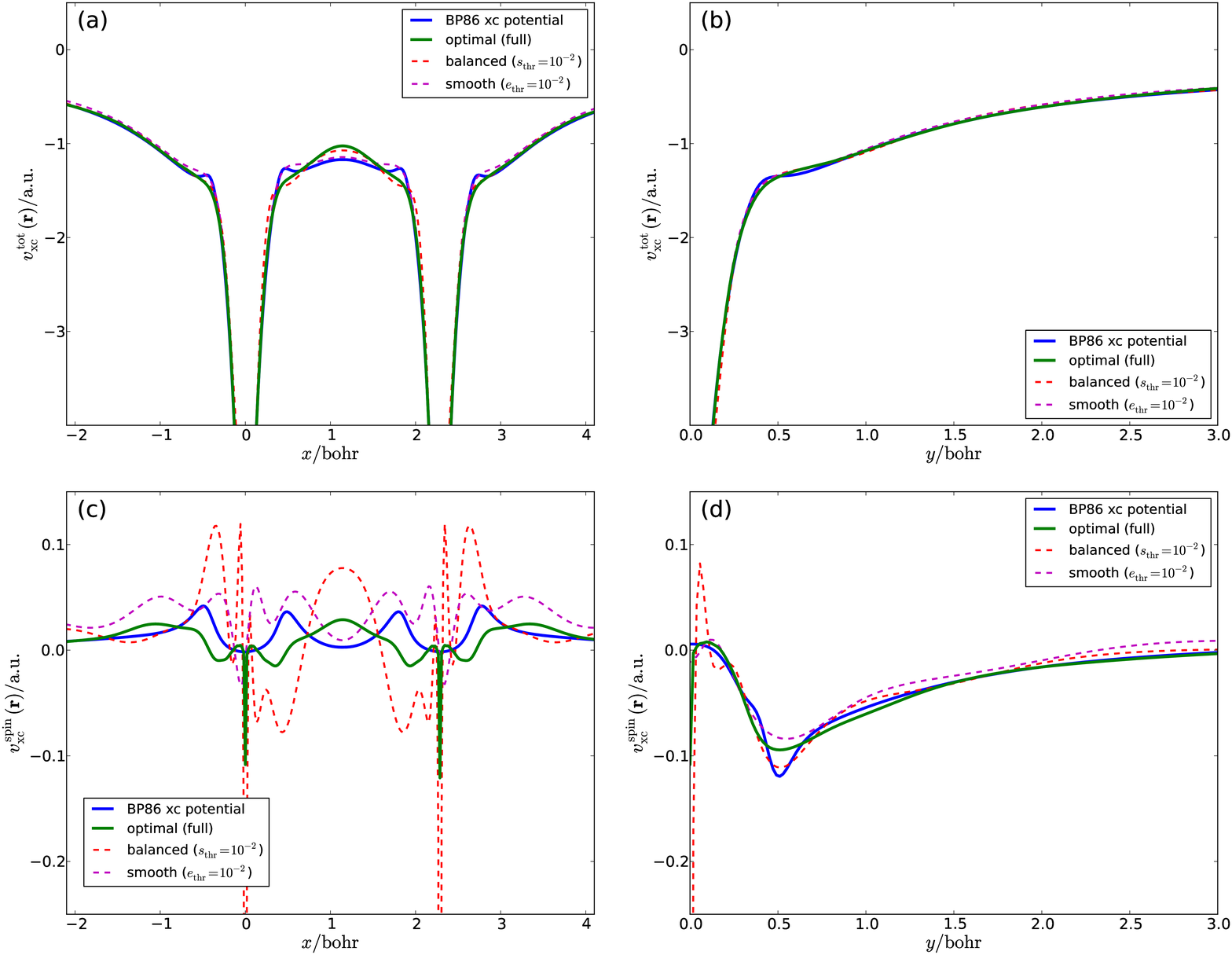}
\end{figure}

The total exchange--correlation potentials reconstructed for the dioxygen molecule from the BP86/QZ4P target density
using the different scheme for singling out an unambiguous potential are shown in Fig.~\ref{fig:o2potentials-bp}a and~b 
along the bond axis and perpendicular to it, respectively. 
For comparison, the figures include the BP86 exchange--correlation potential calculated from 
the target density, i.e., the exchange--correlation potential used for determining the target density. Note, however, that 
because the finite QZ4P basis set is employed when calculating the target density, this BP86 potential is not equal to the
exact potential corresponding to the target density \cite{jacob2011,de_silva_exact_2012}. 

The different reconstructed total exchange--correlation potentials all agree rather accurately with the BP86 exchange--correlation
potential on the scale of the plots. Recognizable differences are only observed along the bond axis in the region between the
oxygen atoms. The BP86 potential shows a plateau in this bond region, which is also reproduced when singling out the
optimized potential that is as smooth as possible. On the other hand, the balanced as well as the optimal potential do not exhibit
such a plateau, but have a maximum at the midbond point. Similar observations can be made for the individual $\alpha$- and 
$\beta$-electron potentials (see Fig.~7 in the Supporting Information).

\begin{table}[h]
\caption{
Absolute errors $\Delta^{\sigma, \rm finite}_{\rm abs}$ in the $\alpha$- and $\beta$-electron densities in the finite orbital
basis set with respect to the 
target $\alpha$- and $\beta$-electron densities (in ${\rm e\, bohr}^{-3}$) obtained with different reconstructed potentials for the 
dioxygen molecule. Results are shown for both the target densities from BP86 and from a CAS(12,12)SCF
calculation and in the latter case using both the TZ2P and with the QZ4P orbital basis set in the potential 
reconstruction.}
\label{tab:error_o2}
{\small
\begin{center}
\begin{tabular}{l cc c cc c cc}\hline \hline
     &  \multicolumn{2}{c}{BP86/QZ4P} & & \multicolumn{2}{c}{CASSCF/TZ2P} & & \multicolumn{2}{c}{CASSCF/QZ4P}  \\
     \cline{2-3} \cline{5-6} \cline{8-9}
     &  $\Delta_{\rm abs}^{\alpha,\rm finite}$ &  $\Delta_{\rm abs}^{\beta,\rm finite}$  & 
     &  $\Delta_{\rm abs}^{\alpha,\rm finite}$ &  $\Delta_{\rm abs}^{\beta,\rm finite}$  & 
     & $\Delta_{\rm abs}^{\alpha,\rm finite}$ & $\Delta_{\rm abs}^{\beta,\rm finite}$ 
     \\ \hline 
Wu--Yang                                   			& 0.0002 & 0.0002 &		& 0.0419 & 0.0439 & 	& 0.0412  & 0.0268    \\
Balanced ($s_{\rm thr} = 10^{-2}$)      	& 0.0019 & 0.0010 &		& 0.0521 & 0.0537 & 	& 0.0414  & 0.0278    \\
Smooth ($e_{\rm thr} = 10^{-2}$)		& 0.0077 & 0.0078 &		& 0.0426 & 0.0462 & 	& 0.0416  & 0.0277    \\
Optimal (full)                             			& 0.0284 & 0.0230 &		& 0.0976 & 0.0830 & 	& 0.0640  & 0.0509    \\ 
\hline
\hline
\end{tabular}
\end{center}
}
\end{table}

The absolute errors $\Delta^{\sigma, \rm finite}_{\rm abs}$  in the $\alpha$- and $\beta$-electron densities 
obtained from the different reconstructed potentials within
the finite orbital basis set compared to the target densities are listed in Table~\ref{tab:error_o2}. Since the QZ4P orbital
basis set used for representing the target density is also employed in the potential reconstruction, the Wu--Yang optimization
can be converged such that the error in the $\alpha$- and $\beta$-electron densities is below $2.0 \cdot 10^{-4}$~${\rm e\, bohr}^{-3}$.
This error increases by one order of magnitude for the potential obtained with an implicitly balanced potential basis set and
further increases when determining the optimized potential that is as smooth as possible. Note that these errors depend on
the choice of the threshold for discarding small singular values $s_{\rm thr}$ and for the change in the density $e_{\rm thr}$,
respectively. For the optimal potential, the finite-basis set absolute error in the $\alpha$- and $\beta$-electron densities
increases further to $2.8 \cdot 10^{-2}$ and $2.3 \cdot 10^{-2}$~${\rm e\, bohr}^{-3}$, respectively. The differences between
the reconstructed total and the spin densities and the respective target (spin) density is illustrated in Fig.~14 in the 
Supporting Information. This comparison shows that except for the region close to the nuclei, all reconstructed potentials
reproduce all qualitative features of the target spin density.

However, to judge the quality of the different potentials, it would be necessary to determine the absolute errors in the
$\alpha$- and $\beta$-electron densities obtained from these potentials in a numerical solution of the KS equations.
Unfortunately, this is not easily possible for the molecular system considered here. Nevertheless, for the lithium atom
considered above and the atomic systems investigated in Ref.~\cite{jacob2011}, it was demonstrated that this error
is the smallest for the optimal potential. Thus, we expect that also for the dioxygen molecule, the optimal potential should
be closest to the exact potential, and that the increased absolute errors in the finite basis set arises because it is not 
possible to reproduce the target density both in a numerical calculation and in the finite QZ4P orbital basis set. Note that
the criterion for singling out the optimal potential [Eq.~\eqref{deltarhofullmin}] minimizes the error in the numerical densities, 
whereas the other approaches minimize the finite-basis set error.

Finally, we turn to the reconstructed spin exchange--correlation potentials, which are presented in Figs.~\ref{fig:o2potentials-bp}c and~d.
As for the lithium atom, reconstructing this spin potential is more difficult than reconstructing the total potential, because it is determined 
as the difference between the $\alpha$- and $\beta$-electron potentials and its accurate determination thus requires that these are both 
obtained with uniform accuracy.
In the plots along the bond axis (see Fig.~\ref{fig:o2potentials-bp}c), both the smooth and the balanced spin potentials deviate significantly 
from the BP86 spin potential. In particular the latter shows rather pronounced oscillations. Note that these oscillations were not visible in
the plots of the individual $\alpha$- and $\beta$-electron potentials, but they are amplified when considering the spin potential. 
The optimal spin potential plotted along the bond axis shows a more regular form and no oscillations that would appear unphysical, but
it also differs from the BP86 spin potential. Nevertheless, except for the spikes at the nuclei themselves, it has a similar overall shape 
near the nuclei, i.e., a symmetric well in which the potential is smaller at the nucleus than ca.~0.5~bohr away from it. This shape of the
spin potential corresponds to the positive spin density close to the nuclei. However, in the midbond region, the shape of the optimal spin 
potential qualitatively differs from the BP86 spin potential and the optimal spin potential  shows a maximum at the midbond point, 
whereas the BP86 spin potential has a minimum. Here, the maximum of the optimal spin  potential is actually in line with the negative 
spin density at the midbond point. 
Despite the observed differences, it appears that of the different reconstructed spin potentials, the optimal potential is closest to the 
BP86 spin potential. Note again that the BP86 spin potential differs from the exact spin potential, which is not available to us here. 
Furthermore, it is important to recall that the singly occupied orbitals have a node at the bond axis, i.e., the spin potential here is only 
responsible for the rather small spin polarization of the doubly occupied orbitals (see Fig.~\ref{fig:o2-dens}a) and, therefore, it should 
be relatively small itself.

\begin{figure}[h]
\caption{(a) Reconstructed spin exchange--correlation potential $v_\text{xc}^\text{spin}$ determined for the dioxygen molecule 
and a BP86/QZ4P target (spin) density in the $xy$-plane. Here, only the optimal potential reconstructed within a QZ4P orbital 
basis set is included. For comparison, (b) shows the BP86 exchange--correlation potential calculated from the target density.}
\label{fig:o2potentials-bp-2d}
\vspace{3ex}\hspace*{-1cm}
\includegraphics[width=1.15\linewidth]{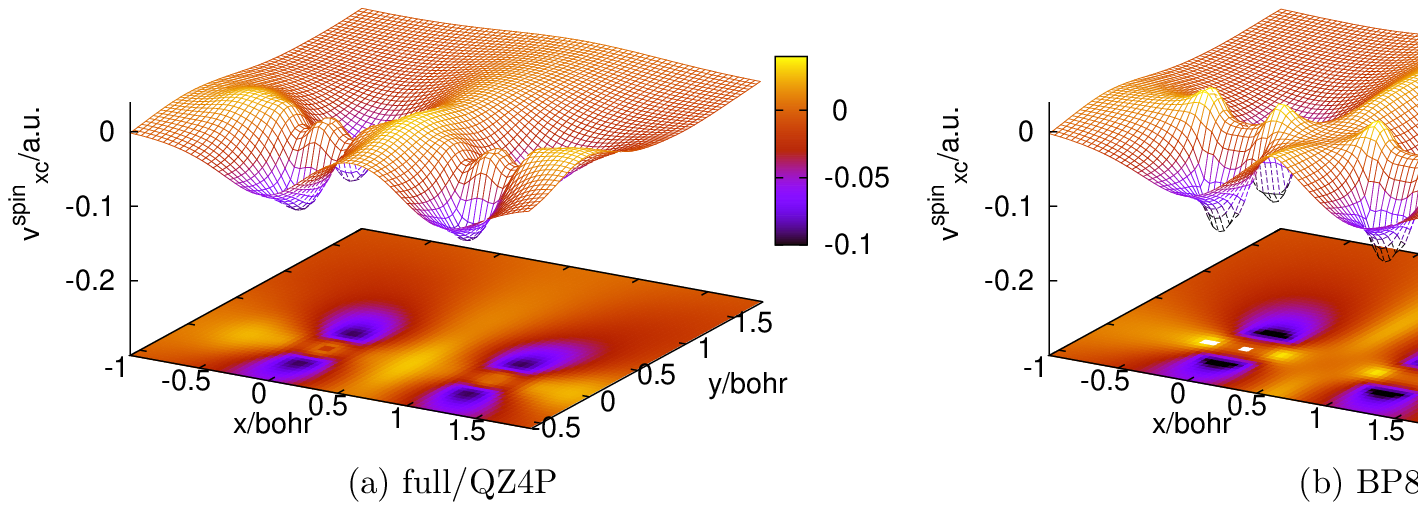}
\end{figure}

When inspecting the reconstructed spin potentials perpendicular to the bond axis (see Fig.~\ref{fig:o2potentials-bp}d), all
reconstructed spin potentials qualitatively agree with the BP86 spin potential. Here, the spin potential is significantly larger
than along the bond axis, as it now covers the region to which the singly occupied orbitals extend (cf.~Fig.~\ref{fig:o2-dens}b).
While the differences are small, the optimal spin potential is closest to the BP86 spin potential and only differs at the nucleus,
whereas both the smooth and the balanced spin potentials show some oscillatory behavior. The overall shape of the optimal
and the BP86 spin potentials in the $xy$-plane is compared in Fig.~\ref{fig:o2potentials-bp-2d}. Except for the differences
along the bond axis, in particular in the midbond region, already discussed above, we observe a good overall agreement.
With the results obtained for the lithium atom in mind, we attribute these differences to the fact that the BP86 potential does
not agree with the exact spin potential, and expect that the optimal spin potential is actually a closer approximation to the
exact spin potential.

\subsubsection*{CASSCF (spin) density expanded in GTOs}

After considering the reconstruction of the spin exchange--correlation potential for the dioxygen molecule for a BP86/QZ4P
target (spin) density, we now turn to an accurate target (spin) density obtained from a CAS(12,12)SCF/cc-pVTZ
calculation. This \textit{ab initio} spin density is compared to the BP86 spin density in Fig.~\ref{fig:o2-dens}. Perpendicular to 
the bond axis (see Fig.~\ref{fig:o2-dens}b), where the spin density is dominated by the unpaired $\pi^*$-electrons, the CASSCF
spin density qualitatively agrees with the one obtained with BP86, even though the magnitude at the maximum and the asymptotic
form slightly differ. Larger differences are found along the bond axis (see Fig.~\ref{fig:o2-dens}a), where the spin density is solely 
due to the spin polarization of doubly occupied orbitals. The CASSCF spin density is significantly larger close to the nuclei, whereas 
the negative spin density at the ends of the molecule is reduced. This is also visible in the plot of the CASSCF spin density in the 
$xy$-plane shown in Fig.~\ref{fig:o2-dens}d.

\begin{figure}[h]
\caption{Reconstructed potentials determined for the dioxygen molecule and a CAS(12,12)SCF/cc-pVTZ target (spin) density. The upper 
part shows the total exchange--correlation potential $v_\text{xc}^\text{tot}$ (a) along the bond axis ($x$ axis) and (b) perpendicular
to the bond axis along the $y$ axis. The lower part shows the spin exchange--correlation potential $v_\text{xc}^\text{spin}$ 
along the (c) $x$ axis and (d) $y$ axis. In the potential reconstruction, the finite QZ4P orbital basis set was employed.
The corresponding plots of the individual $\alpha$- and $\beta$-electron potentials as well as the results obtained with the
TZ2P orbital basis set are shown in Fig.~8 and Figs.~10 and~11 in the Supporting Information, respectively.}
\label{fig:o2potentials-cas}
\vspace{3ex}\hspace*{-1cm}
\includegraphics[width=1.15\linewidth]{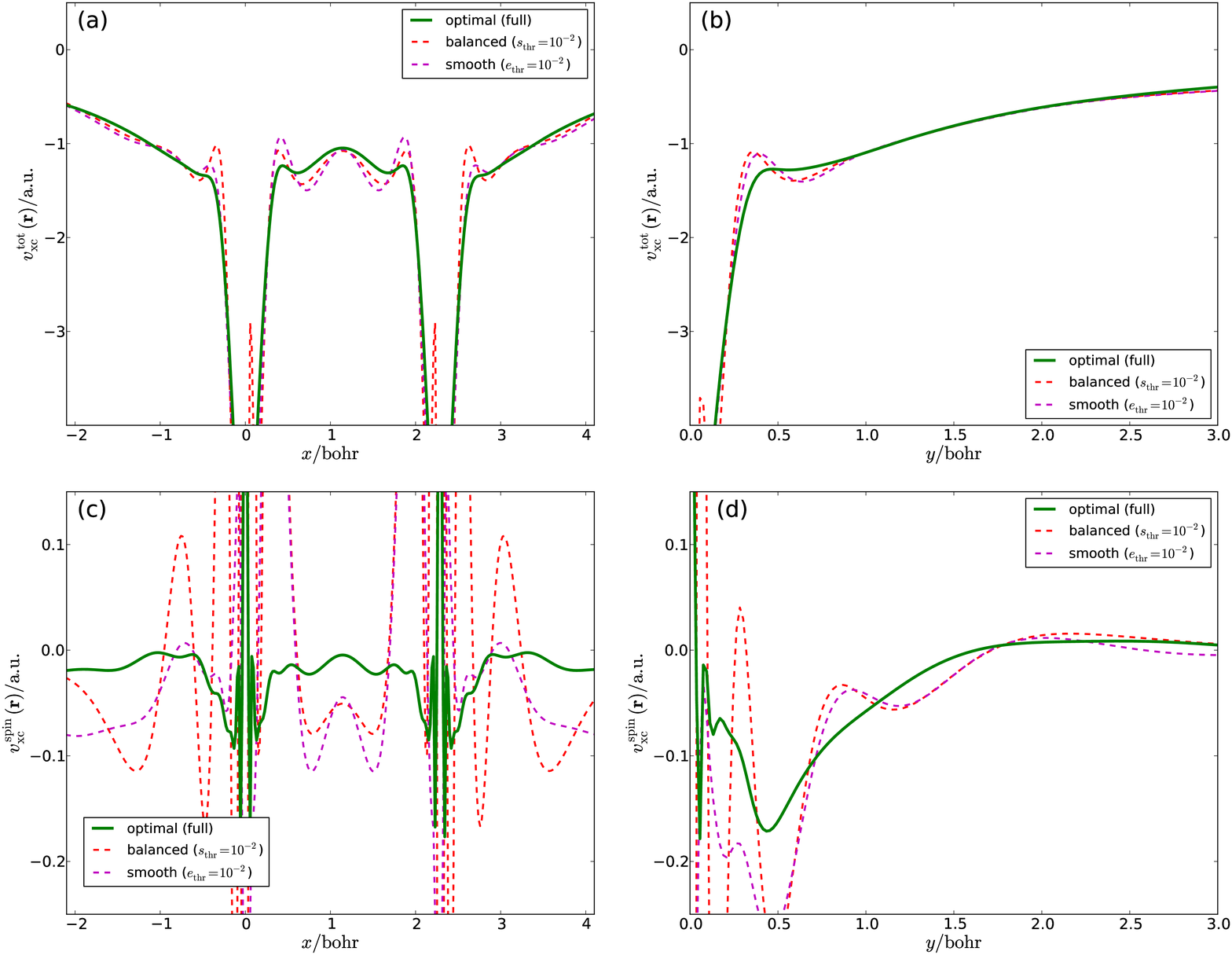}
\end{figure}

While for the BP86/QZ4P target density the BP86 exchange--correlation (spin) potential calculated from the target density 
could be used for comparison --- even though it is not identical to the exact (spin) potential --- we now have no reference 
potential available. Nevertheless, we can still use the results obtained for the BP86 target density to judge whether the
(spin) potentials reconstructed from the CASSCF target (spin) density are physically reasonable. The reconstructed 
total and spin exchange--correlation potentials obtained using a QZ4P orbital basis set using the different scheme for
singling out an unambiguous optimized potential are presented in Fig.~\ref{fig:o2potentials-cas}. The corresponding 
individual $\alpha$- and $\beta$-electron potentials are given in Fig.~8 in the Supporting Information. 

For the reconstructed total exchange--correlation potential, the optimal potential has a similar shape as the BP86 potential
discussed above. Along the bond axis (see Fig.~\ref{fig:o2potentials-cas}a), it has a maximum at the midbond point and exhibits 
some shell structure at ca.~0.5~bohr from the two nuclei. A similar shape as in the outer region along the bond axis is found 
perpendicular to the bond axis for the optimal potential (see Fig.~\ref{fig:o2potentials-cas}). Both the smooth and the balanced 
total potentials have a similar shape, but the shell structure is more pronounced. For the balanced total potential, larger oscillations 
appear close to the nuclei. Nevertheless, the different reconstructed potentials are qualitatively rather similar. A different picture
is obtained if the potential reconstruction is performed with the smaller TZ2P orbital basis set, as is shown in Fig.~11 in the
Supporting Information. In this case, both the balanced and the smooth total potentials show large oscillations, even though
the same thresholds as for the QZ4P orbital basis set are applied. However, the optimal potential is free of such unphysical
oscillations already with the smaller orbital basis set. Moreover, the optimal total potential obtained with the TZ2P orbital basis 
set  is on the scale of the figures in good agreement with the one reconstructed using the larger QZ4P orbital basis set.

The absolute errors in the $\alpha$- and $\beta$-electron densities compared to the target densities obtained from these different 
reconstructed potentials  in the finite TZ2P and QZ4P orbital basis sets are listed in Table~\ref{tab:error_o2}. Because we are now 
trying to reproduce target $\alpha$- and $\beta$-electron densities expanded in a GTO basis set with STOs, these errors are larger 
than for the BP86 target densities that was represented in the same orbital basis also used in the potential reconstruction. With the
TZ2P orbital basis set, the Wu--Yang optimization can be converged until the absolute error is approximately $4 \cdot 10^{-2}$~${\rm e\, bohr}^{-3}$
for both the $\alpha$- and $\beta$-electron density. With the larger QZ4P orbital basis set, a slightly smaller error can be achieved
for the $\alpha$-electron density, whereas the error in the $\beta$-electron density almost halved for \textit{both} electron spins. With both 
orbital basis sets, the errors are larger for the smooth and balanced potentials, with this increase being controlled by the chosen thresholds.
For the larger QZ4P orbital basis set, we observe that the increase is smaller than with the TZ2P orbital basis set. 

The largest density  errors in the finite orbital basis sets are found for the optimal potentials. For the TZ2P orbital basis set, the errors are 
approximately twice as large as for the Wu--Yang potentials. However, a significantly smaller increase is observed with the QZ4P orbital basis 
set. Here, we stress again that for judging the quality of the different potentials, it would be necessary to calculate the errors in the densities
obtained with a numerical solution of the Kohn--Sham equations on the respective potentials. For the optimal potentials, these numerical
density errors, which differ from the finite basis set density errors, should be minimized. With increasing size of the orbital basis set
used in the potential reconstruction, the difference between the finite basis set and the numerical density errors should become smaller. Thus,
the decrease of the finite basis set density errors when going from TZ2P to QZ4P supports our assumption that the optimal potentials 
should be closest to the exact potentials. The total and spin densities calculated from the different reconstructed potentials in the finite 
orbital basis set are compared to the target (spin) density in Fig.~15 in the Supporting Information. Here, it is obvious that these differences 
are larger than for the BP86 target (spin) density. In particular, the optimal potential yields a spin density that misses some of the qualitative
features of the target density around the nuclei. However, it is also apparent that these deviations are reduced when increasing the 
orbital basis set from TZ2P to QZ4P.

In order to study the convergence of the reconstructed potentials with respect to the potential basis set, we performed additional
calculations where additional diffuse functions were included in the potential basis set expansion. The resulting optimal reconstructed 
potentials are shown in Figs.~16--19 in the Supporting Information, both for the TZ2P and the QZ4P orbital basis set. These results show
that the reconstructed potentials are converged with respect to the size of the potential basis set. Only far away from the molecule, i.e.,
in regions where the density is very small, some differences between potentials expanded in different basis sets occur. In addition, we
observe that the requirements on the potential basis set are most severe for the spin potential, where with the TZ2P orbital basis set
increasing the potential basis set still has some effect.

Finally, we now consider the spin exchange--correlation potential,
which is shown in Figs.~\ref{fig:o2potentials-cas}c and~d along the O--O bond axis and perpendicular to it, respectively. Perpendicular
to the bond axis, the optimal spin potential has a similar shape as the one obtained for the BP86/QZ4P target density with a minimum
at ca.~0.5~bohr from the oxygen nuclei, i.e., where the spin density is the largest. In contrast, rather pronounced oscillations are 
observed for the smooth and balanced spin potentials. For the smooth and the balanced spin potentials, such large oscillations are also
present in the plot along the bond axis, whereas the optimal potential is mostly well behaved. Qualitatively, it also resembles the
optimal spin potential reconstructed from the BP86/QZ4P target density. Around the nuclei, it has negative wells, which are deeper
for the CASSCF than for the BP86/QZ4P target density. This is in line with the larger spin density in this region obtained in the
CASSCF calculations. In the bond region between the atoms, the optimal spin potential reconstructed from the CASSCF spin density
is flatter than in the case of the BP86/QZ4P target density, which agrees with the smaller spin polarization in this region. Very
close to the nuclei the reconstructed spin potential has some large oscillations. Most likely, these can be attributed to the 
deficiencies of the GTO target density, which can cause such oscillations in the exact potentials \cite{baerends97}.

\begin{figure}[h]
\caption{Reconstructed spin exchange--correlation potential potentials determined for the dioxygen molecule and a CAS(12,12)SCF/cc-pVTZ 
target (spin) density. For the potential reconstruction with the finite QZ4P orbital basis set (a) the optimal spin potential and (b) the spin potential 
determined by requiring that the optimized potential is smooth are shown. The spin potential obtained by implicitly balancing the orbital 
and potential basis sets as well as the results obtained with the TZ2P orbital basis set are shown in Fig.~9 and Figs~12 and~13 in 
the Supporting Information, respectively.}
\label{fig:o2potentials-cas-2d}
\vspace{3ex}\hspace*{-1cm}
\includegraphics[width=1.15\linewidth]{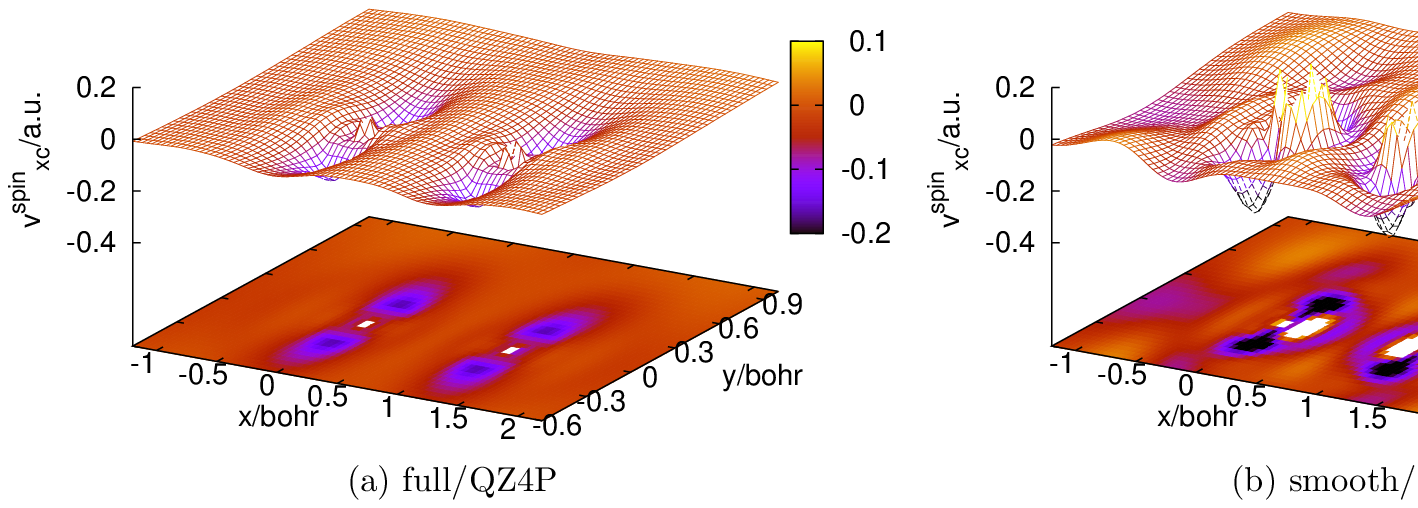}
\end{figure}

To compare the overall performance of the different potential reconstruction schemes, Fig.~\ref{fig:o2potentials-cas-2d} compares the
optimal and the smooth spin potentials in the $xy$-plane. A similar comparison for the balanced spin potential is shown in Fig.~9
in the Supporting Information. While pronounced wiggling features around the position of the nuclei are found in the smooth spin
potential, it is obvious that these can be eliminated in the whole $xy$-plane for the optimal spin potential. Thus, even though there 
is no exact reference spin potential available for comparison, we find that the optimal
spin potential is the only one that seems to be free of artifacts of the potential reconstruction. Moreover, the comparison with the
spin potential reconstructed from the BP86/QZ4P target density shows that the optimal potential is physically reasonable. Therefore,
we are confident that it is the reconstructed potential that is closest to the exact spin potential.

Further support for this conclusion can be drawn from a comparison with the reconstructed potentials obtained for the same CASSCF
target density with the smaller TZ2P orbital basis set, which are shown in Fig.~10 and Fig.~12 and~13 in the Supporting Information. Here,
even larger oscillations are found for the smooth and balanced spin potentials, even though the same thresholds are applied. On the 
other hand, the optimal potentials are qualitatively similar to the QZ4P results. This is particularly obvious for the spin potential perpendicular 
to the bond axis, which is in good agreement with the one reconstructed with the QZ4P orbital basis set. Also along the bond axis
the shapes of the optimal spin potentials are similar for the two orbital basis sets, even though the negative well near the oxygen 
nuclei is less pronounced. Thus, only for the optimal potential a consistent convergence with increasing size of the orbital basis set
is observed.

\section{Conclusions\label{Sec:conclusion}}

In this work, we have extended the unambiguous reconstruction of the local potential yielding a given target 
density \cite{jacob2011} to open-shell systems treated with an unrestricted KS-DFT formalism. Moreover,
we have combined this reconstruction with the use of accurate target (spin) densities obtained from accurate 
wave-function based \emph{ab initio} calculations, i.e., from Full-CI or  CASSCF wave functions. This provides
a route to accurate reference data for the spin exchange--correlation potential $v_{\rm xc}^{\rm spin}$, which 
determines the spin density distribution $Q(\mathbf{r})$ in unrestricted KS-DFT.

Reconstructing this spin exchange--correlation potential is a particularly challenging task, because it is given
by the difference between the reconstructed $\alpha$- and $\beta$-electron potentials. Thus, one has to overcome
the numerical inaccuracies caused by the ill-posed nature of the potential reconstruction problem in finite orbital basis sets, 
as these could otherwise be amplified when calculating the spin potential. As test cases, we chose the lithium atom and the 
oxygen molecule in its triplet state. For both systems, we considered target (spin) densities from unrestricted KS-DFT calculations
as well as from Full-CI (for the lithium atom) and CASSCF (for the dioxygen molecule) calculations. These test cases
made it possible to systematically assess the quality of the reconstructed spin exchange--correlation potentials.

For the lithium atom, it is possible to compare the reconstructed potentials to the exact ones, which can be obtained 
from a fully numerical potential reconstruction. The comparison shows that the optimal spin potentials, determined
using the scheme of Ref.~\cite{jacob2011}, can reproduce the fully numerical spin potential, while the spin potentials
obtained by singling out the $\alpha$- and $\beta$-electron potentials that are as smooth as possible or with an implicitly
balanced potential basis set show significantly larger deviations from the reference spin potential. In general, the quality
of these smooth and balanced spin potentials strongly depends on the choice of the corresponding threshold values and 
can result in highly oscillating potentials if these are chosen too small. On the other hand, if these thresholds are too large, 
the resulting potentials lack all features present in the exact one. Moreover, different thresholds might be required for
reconstructing the $\alpha$- and $\beta$-electron potentials in order to obtain these with similar quality, as is required
for the reconstruction of the spin potential. 

For target densities obtained with GTO orbital basis sets, which are commonly used in wave-function based \textit{ab initio}
calculations, we reconstruct the potential using an STO orbital basis set. This has the advantage that the density obtained
from the Wu--Yang optimization, which is the first step in all potential reconstruction schemes tested here, has the correct
shape close to the nuclei and in the asymptotic region.
Consequently, oscillations close to the nucleus found in the fully numerical reference potential because of the wrong
form the the GTO target density \cite{baerends97} are largely suppressed in the optimal reconstructed potentials.
On the other hand, it requires the use of sufficiently large STO orbital basis sets in order to be able to reproduce the
\textit{ab initio} target (spin) density. For the lithium atom, we found that the a QZ4P orbital basis set is sufficient
to obtain a good agreement with the target density. Moreover, we note that technical improvements to our implementation,
in particular the use of a Tikhonov regularization in the Wu--Yang optimization \cite{Wu2003b} and the use of a cut-off value
for discarding small density regions in the criterion of Eq.~\eqref{deltarhofullmin}, were necessary to treat GTO target densities. 

For the dioxygen molecule, a direct comparison to a fully numerical reference spin potential is not possible. Nevertheless,
for the target density obtained from an unrestricted KS-DFT calculation the exchange--correlation potential used for
determining the target potential can provide some guidance, even though it differs from the exact potential
because of the use of a finite orbital basis set. This comparison shows that the optimal reconstructed spin potential (i.e., the 
one obtained using the scheme of Ref.~\cite{jacob2011}) shows the best overall agreement. Also for the CASSCF target
density, this scheme is the only one that provides a physically reasonable spin exchange--correlation potential, while
the smooth and the balanced potentials are plagued by unphysical oscillations. The optimal potential shows such
oscillations only very close to the nuclei, where they are probably due to deficiencies of the GTO expansion used
for the target density, and to a smaller extent far away from the molecule. 

In summary, we believe that the potential reconstruction scheme proposed in Ref.~\cite{jacob2011} and extended
here to open-shell systems provides the first reliable approach for reconstructing the spin exchange--correlation
potential from accurate \textit{ab initio} (spin) densities. The availability of such accurate reference spin potentials 
can facilitate the development of improved spin-dependent exchange--correlation density functionals which apart from 
yielding accurate total electron densities also provide reliable spin densities.  Thus, this work  represents a prerequisite 
for the design of approximate exchange--correlation functional with an improved spin-density dependance.

\section*{Acknowledgments}

M.R.\ and K.B.\ gratefully acknowledge financial support by a grant from the Swiss national science 
foundation (SNF). K.B.\ thanks the Fonds der Chemischen Industrie for a Chemiefonds scholarship.
C.R.J. acknowledges funding from the DFG-Center for Functional Nanostructures (CFN) at KIT.

\end{document}